\documentclass[11pt,a4paper]{article}
\usepackage{jcappub}
\usepackage{graphicx}% Include figure files
\usepackage{dcolumn}% Align table columns on decimal point
\usepackage{amsfonts}
\usepackage{amssymb}
\usepackage{bm}% bold math
\usepackage{hyperref}
\usepackage{url}
\usepackage{subfigure}
\usepackage{natbib}
\usepackage{txfonts}
\usepackage{color}

\begin{document}

\newcommand{\changeR}[1]{\textcolor{red}{#1}}
\newcommand{\changeRII}[1]{\textcolor{red}{#1}}
\newcommand{\DTe}{{{\Delta_{T_e}}}}
\newcommand{\SQ}{{{\mathcal{E}}}}
\newcommand{\TBB}{{{T_{\rm BB}}}}
\newcommand{\TBE}{{{T_{\rm BE}}}}
\newcommand{\TCMB}{{{T_{\rm CMB}}}}
\newcommand{\Te}{{{T_{\rm e}}}}
\newcommand{\Teq}{{{T^{\rm eq}_{\rm e}}}}
\newcommand{\Ti}{{{T_{\rm i}}}}
\newcommand{\nB}{{{n_{\rm B}}}}
\newcommand{\nHe}{{{n_{\rm ^4He}}}}
\newcommand{\nH}{{{n_{\rm H}}}}
\newcommand{\nHet}{{{n_{\rm ^3He}}}}
\newcommand{\nHt}{{{n_{\rm { }^3H}}}}
\newcommand{\nHtw}{{{n_{\rm { }^2H}}}}
\newcommand{\nBes}{{{n_{\rm { }^7Be}}}}
\newcommand{\nBE}{{{n_{\rm BE}}}}
\newcommand{\Bes}{{{{\rm { }^7Be}}}}
\newcommand{\nLis}{{{n_{\rm { }^7Li}}}}
\newcommand{\nLisi}{{{n_{\rm { }^6Li}}}}
\newcommand{\nS}{{{n_{\rm S}}}}
\newcommand{\nSS}{{{n_{\rm ss}}}}
\newcommand{\Teff}{{{T_{\rm eff}}}}
\newcommand{\Mpc}{{{{\rm ~Mpc}}}}

\newcommand{\id}{{{\rm d}}}
\newcommand{\aR}{{{a_{\rm R}}}}
\newcommand{\bR}{{{b_{\rm R}}}}
\newcommand{\neb}{{{n_{\rm eb}}}}
\newcommand{\neql}{{{n_{\rm eq}}}}
\newcommand{\kB}{{{k_{\rm B}}}}
\newcommand{\EB}{{{E_{\rm B}}}}
\newcommand{\zmin}{{{z_{\rm min}}}}
\newcommand{\zmax}{{{z_{\rm max}}}}
\newcommand{\YBEC}{{{Y_{\rm BEC}}}}
\newcommand{\yg}{{{y_{\rm \gamma}}}}
\newcommand{\y}{{{y}}}
\newcommand{\rhob}{{{\rho_{\rm b}}}}
\newcommand{\Ne}{{{n_{\rm e}}}}
\newcommand{\sigT}{{{\sigma_{\rm T}}}}
\newcommand{\me}{{{m_{\rm e}}}}
\newcommand{\npl}{{{n_{\rm pl}}}}
\newcommand{\nY}{{{n_{\rm Y}}}}

\newcommand{\kD}{{{{k_{\rm D}}}}}
\newcommand{\KC}{{{{K_{\rm C}}}}}
\newcommand{\KdC}{{{{K_{\rm dC}}}}}
\newcommand{\Kbr}{{{{K_{\rm br}}}}}
\newcommand{\zdC}{{{{z_{\rm dC}}}}}
\newcommand{\zbr}{{{{z_{\rm br}}}}}
\newcommand{\aC}{{{{a_{\rm C}}}}}
\newcommand{\adC}{{{{a_{\rm dC}}}}}
\newcommand{\abr}{{{{a_{\rm br}}}}}
\newcommand{\gdC}{{{{g_{\rm dC}}}}}
\newcommand{\gbr}{{{{g_{\rm br}}}}}
\newcommand{\gff}{{{{g_{\rm ff}}}}}
\newcommand{\xe}{{{{x_{\rm e}}}}}
\newcommand{\alphafs}{{{{\alpha_{\rm fs}}}}}
\newcommand{\YHe}{{{{Y_{\rm He}}}}}
\newcommand{\SE}{{{\dot{{Q}}}}}
\newcommand{\SN}{{\dot{{N}}}}
\newcommand{\muc}{{{{\mu_{\rm c}}}}}
\newcommand{\xc}{{{{x_{\rm c}}}}}
\newcommand{\xH}{{{{x_{\rm H}}}}}
\newcommand{\mT}{{{{\mathcal{T}}}}}
\newcommand{\Ob}{{{{\Omega_{\rm b}}}}}
\newcommand{\Or}{{{{\Omega_{\rm r}}}}}
\newcommand{\Odm}{{{{\Omega_{\rm dm}}}}}
\newcommand{\mdm}{{{{m_{\rm WIMP}}}}}

\title{Beyond $\y$ and $\mu$: the shape of the CMB spectral distortions 
  in the intermediate epoch, $1.5\times 10^4\lesssim z \lesssim 2\times 10^5$}

%\author{Rishi Khatri\inst{\ref{inst1}}
%\and
%Rashid A. Sunyaev\inst{\ref{inst1}\ref{inst2}}
%}

%\institute{ Max Planck Institut f\"{u}r Astrophysik, Karl-Schwarzschild-Str. 1,
%  85741, Garching, Germany\\ \email{khatri@mpa-garching.mpg.de}\label{inst1}
%\and
% Space Research Institute, Russian Academy of Sciences, Profsoyuznaya
% 84/32, 117997 Moscow, Russia \label{inst2}}
\author[a]{Rishi Khatri,}
\author[a,b,c]{Rashid A. Sunyaev}

\affiliation[a]{ Max Planck Institut f\"{u}r Astrophysik\\, Karl-Schwarzschild-Str. 1
  85741, Garching, Germany }
\affiliation[b]{Space Research Institute, Russian Academy of Sciences, Profsoyuznaya
 84/32, 117997 Moscow, Russia}
\affiliation[c]{Institute for Advanced Study, Einstein Drive, Princeton, New Jersey 08540, USA}
\emailAdd{khatri@mpa-garching.mpg.de}

\date{\today}

\abstract
{We  calculate  numerical solutions and analytic approximations for the
  intermediate-type spectral distortions.  
Detection of a $\mu$-type  distortion (saturated comptonization) in the CMB  will constrain the
  time of energy injection to be at a redshift $2\times 10^6\gtrsim z \gtrsim 2\times 10^5$, while a
  detection of a $\y$-type distortion (minimal comptonization) will mean that there was heating of
  CMB at redshift $z\lesssim 1.5\times 10^4$. We point out that the
  partially comptonized 
  spectral distortions, generated  in the redshift range
  $1.5\times 10^4\lesssim z \lesssim 2\times 10^5$, are much richer in
  information than the pure $\y$ and $\mu$-type distortions.  The spectrum created
  during this period is intermediate between $\y$ and $\mu$-type
  distortions and depends sensitively on the redshift of energy injection. These intermediate-type distortions cannot be mimicked by
  a mixture of $\y$ and $\mu$-type distortions at all frequencies and vice
  versa.  The measurement of these intermediate-type  CMB spectral distortions  has the possibility to constrain precisely not only the amount of energy release
  in the early Universe but also the mechanism, for example, particle
  annihilation and Silk damping can be distinguished from particle
  decay. The intermediate-type distortion templates and software
    code using these templates to calculate the CMB spectral distortions for user-defined energy injection
    rate are made publicly available.
}

%\pacs{06.20.Jr,98.80.-k}
%\titlerunning{Intermediate-type spectral distortions}
\maketitle
\flushbottom

\begin{section}{Introduction}
 Cosmic background explorer (COBE/FIRAS) \citep{cobe}  measurements show
that cosmic microwave background (CMB) follows Planck spectrum to a high
precision between $1<x<11$, {where $x=h\nu/\kB T$ is the dimensionless
frequency, $h$ is Planck's constant, $\kB$ is Boltzmann's constant and $T$
is the temperature of the CMB blackbody spectrum.} The
precision is quantified by the $2\sigma$ limits on the
chemical potential \cite{sz1970} $\mu \lesssim 9\times 10^{-5}$  and $\y$-type
distortion \citep{zs1969} $\y\lesssim 1.5\times 10^{-5}$. Technologically an
improvement of more than two orders of magnitude over COBE/FIRAS has been
possible for some time
\citep{fm2002} and proposed future experiment Primordial Inflation Explorer
(PIXIE) \citep{pixie} will be able to measure a $\y$-type distortion of $\y=10^{-8}$ or
$\mu=5\times 10^{-8}$ at
$5\sigma$, a more than three orders of magnitude improvement over COBE/FIRAS.

In case there is energy or photon production at a redshift $z\gtrsim
2\times 10^6$,
the photon production and destruction through bremsstrahlung and double
Compton scattering along with the redistribution of photons in energy via Compton
scattering {on thermal electrons} can establish full thermal equilibrium  \citep{sz1970,dd1982,ks2012} and we get a blackbody
spectrum with a higher temperature. The redshift $z\approx 2\times
  10^6$ therefore defines the boundary of the blackbody photosphere. This happens, for example, when electrons
and positrons, {having higher initial entropy and energy density
  than photons,} annihilate at $z\sim 10^8\backsim 10^9$.  Also, when there is energy
release during primordial nucleosynthesis, the photon spectrum quickly
thermalizes. For redshifts between $2\times 10^5\lesssim z \lesssim 2\times 10^6$,
any heating of CMB gives rise to a Bose-Einstein spectrum or a $\mu$-type
distortion, where $\mu$ is the chemical potential.  For
redshifts $z\lesssim 1.5\times 10^4$, the Compton {redistribution
  of photons over frequencies is too weak}  to
establish the equilibrium Bose-Einstein spectrum and we get a $\y$-type
spectrum. 
For $\mu$-type ($\y$-type) distortions, if detected, we can only  put a
lower (upper) limit to the time of energy release. However, if there is
heating  of CMB in the redshift range $1.5\times 10^4\lesssim z \lesssim
2\times 10^5$, the
spectrum depends sensitively on the time of energy injection and it is thus
possible to put a much more precise constraint on the time of energy release.
 Important epochs in
the history of the early Universe, from point of view of the CMB spectrum,
are depicted schematically in Fig. \ref{cmbfig}. 

\begin{figure}
\resizebox{\hsize}{!}{\includegraphics{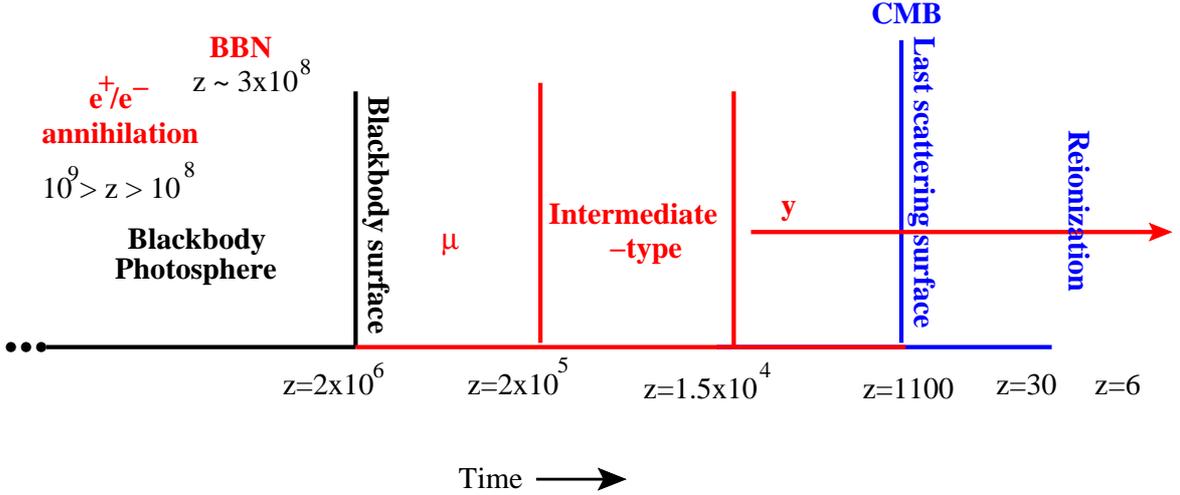}}
\caption{\label{cmbfig} Important epochs related to the creation and
  evolution of  the CMB spectrum.
}
\end{figure}

We first define the spectral distortion as a pure redistribution
  of photons of a reference blackbody,
which is slightly different for $\mu$ type distortions from the
conventional definition. This definition makes it possible to uniquely
define the zero point/crossing frequency of the spectral distortion and
use it to determine the redshift of energy injection. 
We  review the $\y$-type solution and also calculate the regime of
validity of this solution in the Appendices. We  study the comptonization of $\y$-type
distortion and calculate the evolution of zero point/crossing frequency of
the distortion. As a first example, we discuss the shape of the  spectral distortions arising
from the dissipation of {sound waves in the early Universe \citep{silk,sz1970b,daly1991,hss94,cs2011,cks2012,pz2012}} and how they can
be used to constrain the shape of the primordial power spectrum on small
scales with wavenumbers $8 <k<378 \Mpc^{-1}$. Additional significant
sources of heating/cooling in the early Universe include  
dark matter  (WIMP) 
annihilation and adiabatic cooling of baryons \citep[][see also
\citealp{ksc2012,ks2012}]{cs2011}. The $\y$-type distortions are also created throughout the late-time history
of the universe, in particular from the heating of CMB by hotter electrons {in the
intergalactic medium during and after
reionization, and by hot intracluster electrons in the clusters of galaxies.}   The early Universe contributions to the $\y$-type
distortions, before recombination, are indistinguishable from the late-time
contributions; $\y$-type distortions are therefore not very useful in
constraining possible new physics, which can heat the CMB before
reionization. As long as the distortions are small,  the
$\y$-type distortions from different epochs and  the $\mu$ and the
intermediate-type distortions just add linearly to give the total
distortion to the CMB. Evolution of spectral distortions in the
early Universe was first considered by Zeldovich and Sunyaev \citep{zs1969,sz1970} and analytic
and 
numerical solutions were computed by \citep{is1975}. These  and later
calculations 
 \citep{dd1982,bdd1991,hs1993,pb2009,cs2011}, 
which included double Compton scattering and considered low baryon density
Universes such as ours, were focused on $\mu$ and $\y$-type distortions,
although some  authors computed the full evolution of the spectrum,
including intermediate-type contributions. 
We should clarify that in this
  paper we are always in the non-relativistic regime and are not concerned
  about the relativistic corrections to either the Kompaneets equation
  \citep{k1956} or the $\y$-type distortions, which have been studied in some
  detail by various authors
  \citep{wright1979,Rephaeli1995,CL1998,Itoh1998,ss1998,Nozawa1998,MB1999,EK2000,dolgov2001,SR2004,chluba2005,chluba2012} {in applications to the very hot ($\Te\gtrsim 10~\rm{KeV}$) gas inside clusters of galaxies}.

In the following calculations we will use WMAP cosmological parameters of
$\Lambda CDM$ cosmology \citep{wmap7} with CMB temperature today
$\TCMB=2.725$K, number of neutrinos $N_{\nu}=3.046$ \citep{mm2005}, 
helium mass fraction $\y_{He}=0.24$, Hubble constant $H_0=70.2\rm{km/s/Mpc}$,
matter density $\Omega_m=0.275$, baryon density $\Omega_b=0.0458$ and zero curvature.

\section{Possible non-standard sources of energy injection in the early Universe}
Detection of energy release in the early Universe is an important source of
information about new physics. There are several possible theoretical
sources from high energy theories. For example, in super-symmetric theories
and Kaluza-Klein theories it is possible that dark matter was initially
produced as a long lived next to
lightest particle in the dark sector  and then decays
later to the lightest  particle which acts as dark matter today \cite{fengreview}. 
An example is a neutralino  ($\tilde{B}$) decaying to gravitino
($\tilde{G}$) in super-symmetric theories. Neutralino in this case is a WIMP
which decays into gravitino which has only gravitational interaction is
thus 
super weakly interacting or SWIMP \cite{fengsusy}. Sterile neutrinos can
also be
a significant component of dark matter and their decay can be a source of
energy and photons  \cite{fengreview}.
Other sources of energy release include evaporating 
primordial black holes  \cite{pbh},  decaying cosmic strings and other
topological defects, cosmic string wakes \citep{csbook} and
oscillating super-conducting cosmic strings \citep{vilenkin1988,tsv2012}, and
small-scale primordial magnetic fields \cite{Jedamzik}.

In most of the above examples the decay of an initial particle  results in
high energy electromagnetic and hadronic showers. Initial  standard model
particles,  that  are produced as a result of these showers,
 interact with electrons, ions and photons through a rich variety
of energetic processes like Compton and
inverse Compton scattering, pair production, photon-photon interaction
etc. (see for example \cite{spf2009}) depositing
most of the initial energy in the form of heat in plasma very quickly. Some
of the energy is lost to neutrinos and energetic particles in the windows
of low optical depth. This
energy deposition in turn gives rise to a $\y$-type distortion which then
comptonizes. As a second example, we consider decay of an unstable particle
as a source of energy injection and show how the intermediate-type spectral
distortions can constrain the life time of the particle as well as
distinguish it from energy injection due to a process for which the  rate
of 
 energy injection as a function of redshift is not exponential but a power
 law, such as annihilation of particles {or dissipation of sound waves.}

\end{section}

\section{Definition of CMB reference temperature and spectral distortion}
For any given isotropic unpolarized photon spectrum , $n(\nu)$, where $n$ is the occupation number
as a function of frequency $\nu$, we can define a unique reference
temperature. {At redshifts $z\lesssim 2\times 10^6$, the energy
  exchange between the plasma and radiation is very fast, but the
  production (and absorption) of photons is extremely slow. As a result,
  the number density of photons does not change due to energy release and
  the ratio of photon to baryon number density $n_{\gamma}/n_{\rm b}$ is
  constant to high precision.\footnote{{There is an exception, of
      course,  when the energy
  release mechanism also adds photons, for example Silk damping \citep{cks2012,ksc2012b}.}}} We can calculate the total number density  of photons in the
spectrum
\begin{align}
N=\int \frac{8 \pi \nu^2}{c^3}n(\nu)\id \nu,
\end{align}
where $c$ is the speed of light. We can now define a reference temperature, $T$,
as the temperature of the blackbody spectrum $\npl$ which has the same number
density $N$. Thus $T=(N/\bR)^{1/3}$, where $\bR=\frac{16 \pi\kB^3\zeta(3)}{ c^3 h^3}$, $\zeta$ is the Riemann zeta
function with $\zeta(3)\approx 1.20206$. We can now also define dimensionless
frequency, $x=h\nu/(\kB T)$ and write the total spectrum as 
\begin{align}
n(x)=\npl(x)+\Delta n(x),
\end{align}
where $\Delta n$ is the distortion from the reference blackbody with the
property that it represents a redistribution of photons in the reference
blackbody, $\npl=1/(e^x-1)$. The total number of photons in the distortion
vanishes,  $\int\id x ~ x^2 \Delta n \equiv 0$.  This
formalism is especially useful for discussing the distortions created in the early
Universe solely through the action of comptonization, which just
redistributes the photons already existing in a previous spectrum, and thus
does not change the reference temperature defined above. Of
  course, many other definitions of spectral distortions are possible by
  defining a different reference temperature \cite{cs2011,cks2012}.  The above definition is just an extension of the
  conventional definition of a $\y$-type distortion to distortion of any shape.

\begin{figure}
\resizebox{\hsize}{!}{\includegraphics{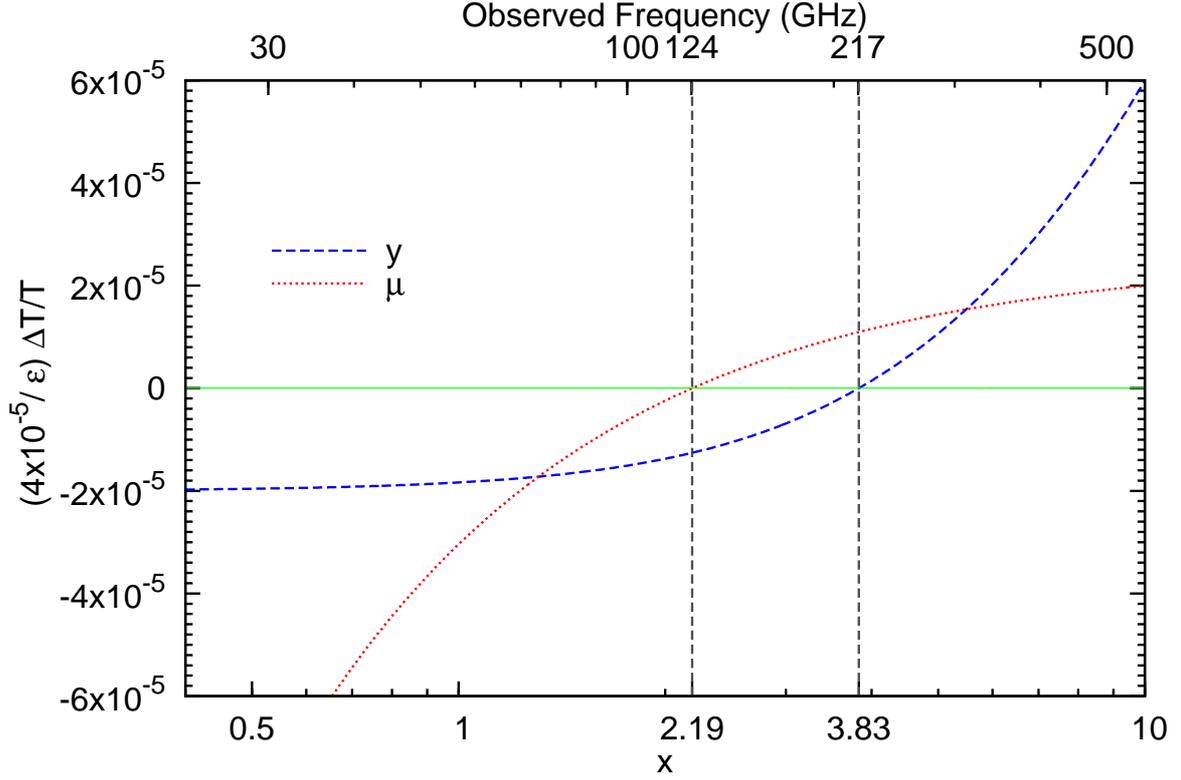}}
\caption{\label{t2fig} $\y$-type and $\mu$-type distortions created by
  addition of energy $\mathcal{E}\equiv\Delta E/E =4\times 10^{-5}$ to a blackbody with
  reference temperature $T$. Fractional difference in effective temperature defined in Eq. \ref{teffeq}
  is plotted.}
\end{figure}

\begin{figure}
\resizebox{\hsize}{!}{\includegraphics{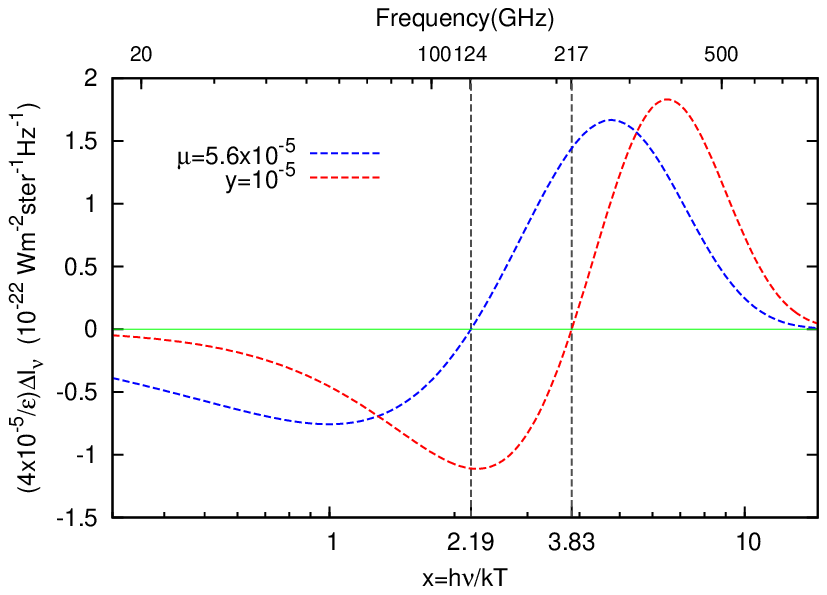}}
\caption{\label{f2fig} Same as Fig. \ref{t2fig} but difference in intensity
from a blackbody with reference temperature $T$ is plotted.}
\end{figure}

In the rest of the paper we will use the above definitions of reference
temperature and spectral distortions. We note that for the case of a
Bose-Einstein spectrum $\nBE$, the above definition gives a different reference
temperature than the one given by the usual definition
$\nBE=1/\left(e^{\frac{h\nu}{\kB T_{\rm BE}}+\mu}-1\right)$. Let us remind
that $\nBE$ is the
equilibrium solution of the Kompaneets equation \citep{k1956} when
$\Te=\TBE$, where $\Te$ is the electron temperature. Our definition
corresponds to  taking an
initial blackbody spectrum with temperature equal to the reference
temperature $T$ and add to it small amount of energy (keeping the photon
number constant), which then fully comptonizes creating a Bose-Einstein
spectrum with a chemical potential $\mu$ fully defined by the amount of
energy release. It is easy to calculate the
final temperature of the resulting spectrum using relations for number
density and energy density of a Bose-Einstein spectrum in the limit of
small chemical potential \citep{is1975b} and it is given by $t\equiv
(\TBE-T)/T=0.456\mu$. Thus the spectrum written in terms of  dimensionless
frequency $x=h\nu/\kB T$ is given by, 
\begin{align}
n_{\rm BE}&=\frac{1}{e^{\frac{h\nu}{\kB T_{\rm
          BE}}+\mu}-1}\nonumber\\
&=\frac{1}{e^{x-0.456\mu x+\mu}-1}\nonumber\\
&\approx \npl(x)+\frac{\mu e^x}{\left(e^{x}-1\right)^2}\left(\frac{x}{2.19}-1\right),
\end{align}
giving the zero point, where the coefficient of $\mu$ in above equation vanishes making the
spectrum identical to the blackbody spectrum at temperature $T$,  at $x_0=2.19, \nu=124 ~{\rm GHz}$, compared to
$x_0=3.83, \nu=217~{\rm GHz}$ for a $\y$-type distortion \citep{zs1969},
{
\begin{align}
n_{\y}(x)=\frac{xe^x}{(e^x-1)^2}\left[x\left(\frac{e^x+1}{e^x-1}\right)-4\right]\label{linsz}.
\end{align}
}
 Finally we note
that part of the energy release from sound wave dissipation in the early
Universe goes into blackbody part of the spectrum \citep{cks2012}.  The
blackbody part of the energy release is easily absorbed in
the reference temperature $T$ and can be ignored completely in our method, where we define
the spectral distortions as pure redistribution of photons of a reference
blackbody.\footnote{If we add of order $\epsilon$ fractional energy to a blackbody of
  initial temperature $T$ with part of it going into blackbody and part into
  $\mu$ distortion, then  for the new temperature $T'$ we have,
  $(T'-T)/T\sim \mu\sim \epsilon$. Ignoring the energy addition to the
  blackbody part we will use
  the frequency variable $x$, while the frequency variable with respect to
  the new
  reference temperature is $x'=h\nu/(\kB T')$. The error introduced in the
  calculation of distortions is thus second order in $\epsilon$,
  $\Delta n(x)/n(x)=\Delta n(x')/n(x')+\mathcal{O}(\epsilon^2)$, and can be
ignored for small distortions. In addition, this change in blackbody
temperature takes no part in comptonization, since it also changes the electron
temperature by the same amount (Eq. \eqref{te}). }
We have plotted in  Fig\ref{t2fig}.  $\y$-type and $\mu$-type (Bose-Einstein) distortions
having same number and energy density.  The fractional difference in the
 effective temperature is plotted, which is  defined by the following equations.
\begin{align}
n(x)&\equiv \frac{1}{e^{h \nu/k_B \Teff(x)}-1}\nonumber\\
    &=\frac{1}{e^{x T/ \Teff(x)}-1}\nonumber\\
\Rightarrow \frac{\Delta T}{T}& \equiv \frac{\Teff-T}{T}\approx \frac{1-e^{-x}}{x}
  \frac{\Delta n}{\npl}\label{teffeq}.
\end{align}
Fig. \ref{f2fig} shows the difference in intensity from blackbody, {$\Delta I_{\nu}\equiv
I_{\nu}^{\mu,\y}-I_{\nu}^{\rm pl}\propto x^3 \Delta n$, where $I_{\nu}^{\mu,\y}$ is the intensity
of $\mu$ or $\y$ distorted spectrum, and $I_{\nu}^{\rm pl}$ is the intensity
of Planck spectrum at reference temperature $T$ defined above. The two type of
distortions shown ($\mu$ and $\y$)  have the same energy density.}

\begin{section}{Creation of $\y$-type distortion, role of Compton $\yg$
    parameter and validity of the solution}
 A source of energy injection,
for example dark matter decay, will in general lead to a shower of
particles which will quickly deposit most of their energy in the plasma,
and result in an increase in the electron
temperature as long as the source is on. We will first review the  $\y$-type
distortions created as a result  of a source of energy that turns on for a very
short time and also discussed the regime of validity of this solution.

\begin{figure}
\resizebox{\hsize}{!}{\includegraphics{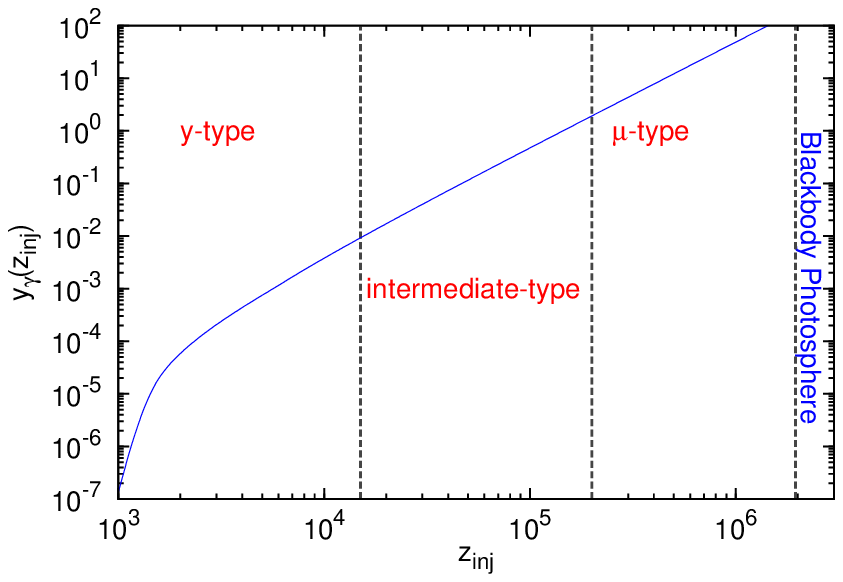}}
\caption{\label{yfig}Dependence of Compton parameter on redshift, $\yg(z_{\rm{inj}})=-\int_{z_{\rm{inj}}}^{0}dz\left[k_B\sigma_Tn_eT\right]/\left[m_e
  cH(1+z)\right]$. The drop in the plot at  $z\sim 1000$ is due to the depletion of
electrons because of recombination.  The total energy released is divided
approximately equally between $\y$-type, intermediate-type  and $\mu$-type
distortions for the case of Silk damping with spectral index of initial
power spectrum close to unity \citep[e.g. see][]{ksc2012} and for other
mechanisms with similar rate of energy injection, e.g.  dark matter
annihilation. For energy release from exponential decay of particles, the
division of energy is less democratic and depends sensitively on the
lifetime of the particle. {Recombination \citep{zks68,peebles68,sss2000} was
calculated using the effective multilevel approach \citep{ah2010} following
publicly available codes HyRec \citep{ah2011} and CosmoRec \citep{ct2011}.}}
\end{figure}

\subsection{Compton parameter $\yg$ and the Compton distortion parameter $\y$.}
The equation describing the evolution of photon spectrum through Compton
scattering is the Kompaneets equation \citep{k1956}. We will work with the dimensionless
frequency $x$ and the photon occupation number $n(x)$ which is given in the
case of blackbody spectrum at temperature $T=2.725 (1+z)~{\rm K}$ by
$\npl(x)=1/(e^x-1)$. In the case of a blackbody or a Bose-Einstein
spectrum at a different temperature, there will be  factors associated
with the rescaling of temperature. Working with $n(x)$, which is invariant
with respect to the adiabatic expansion of photon gas,  allows us to  avoid the extra terms in the equations
associated with the expansion of the Universe. To simplify equations
further,  we will use as our time variable the
parameter $\yg$ (not to be confused with the $\y$-type distortion  parameter $\y$) defined as follows:
\begin{align}
\yg(z,z_{\rm{max}})=-\int_{z_{\rm{max}}}^{z}dz\frac{k_B\sigma_T}{m_e
  c}\frac{n_eT}{H(1+z)},\label{yz}
\end{align}
where $z_{\rm{max}}$ is the redshift where we start the evolution of the
spectrum or the energy injection redshift. It is convenient to use
radiation temperature $T$ and not electron temperature
$\Te$ in the definition of $x$ and $\yg$, as it
makes $x$ independent of the expansion of the Universe after
electron-positron annihilation, with both $T, \nu \propto (1+z)$. The electron
temperature on the other hand  evolves in a non-trivial way after baryons
thermally decouple from radiation at $z\sim 500$.  The total $\yg$ parameter for an
energy injection redshift of $z_{\rm{inj}}$ is then given by
$\yg(z_{\rm{inj}})\equiv \yg(0,z_{\rm{inj}})$.
 The
Compton parameter $\yg(z_{\rm{inj}})$ is plotted in Fig. \ref{yfig}  and it can be seen
that the contribution to the integral from $z<1.5\times 10^4$ becomes very
small, with $\yg
\lesssim 0.01$.  We will present our results as functions of  $\yg$ and they can be
converted into functions of redshifts  using Eq. \ref{yz} and
Fig. \ref{yfig}.
During radiation domination, we can calculate the $\yg$ parameter analytically
and is given by $\yg(z,z_{\rm max})\approx \frac{k_B\sigma_T}{m_e
  c}\frac{(\nH+2\nHe) \TCMB}{2
  H_0\Omega_r^{1/2}}\left[(1+\zmax)^2-(1+z)^2\right]=4.88\times
10^{-11}\left[(1+\zmax)^2-(1+z)^2\right],$ where $\nH$ and $\nHe$ are the
number densities of hydrogen and helium nuclei today respectively, and
$\Or$ is the radiation energy density today in units of critical density.
{Similarly, it is easy to find analytic formulae during
the matter dominated era, $z\ll 3200$, but before recombination and also after
reionization, when the
electron density is again simply equal to sum of hydrogen and helium number
densities (assuming singly ionized helium).}

\begin{figure}
\resizebox{\hsize}{!}{\includegraphics{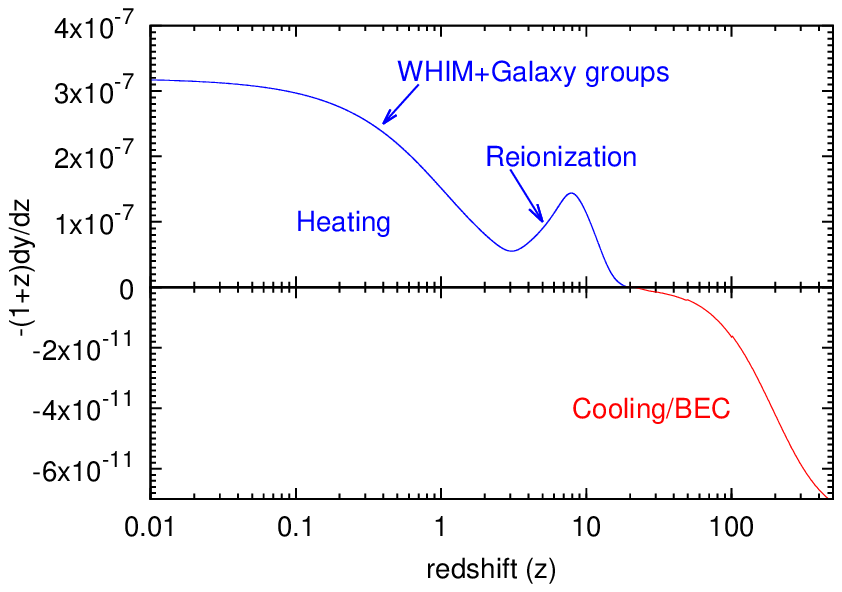}}
\caption{\label{Yrifig}{Post-recombination (rough) estimate of the thermal $\y$-distortion injection rate
  $-(1+z)\id \y/\id z$ is plotted, it is approximately equivalent to the energy
release in the redshift interval $\delta z\sim z$.} Before reionization starts, the
  distortions are dominated by cooling of CMB or Bose-Einstein
  condensation (BEC) due to comptonization with
  colder electrons giving a negative $\y$ distortion. During and after
  reionization the intergalactic medium is heated to temperature
  $\Te\gtrsim 10^4$ giving a much larger positive $\y$ distortion of
  amplitude $\sim 10^{-7}$. {There will also be similar magnitude
 positive  { contributions from the second order Doppler effect
   (not shown above) arising due to peculiar velocities of electrons}.} }
\end{figure}

We also define the Compton $\y$-type distortion amplitude, $\y$,
{
\begin{align}
\y&=\int_0^\yg \DTe \id \yg\equiv -\int_\zmax^z \frac{k_B\sigma_T}{m_e
  c}\frac{n_e(\Te-T)}{H(1+z)}\id z\nonumber\\
&\approx^{\Te\gg T} -\int_\zmax^z \frac{k_B\sigma_T}{m_e
  c}\frac{n_e\Te}{H(1+z)}\id z\equiv y_{\rm e},
\end{align}
}
where, $\DTe\equiv T_e/T-1$, and $T_e$ is the electron
temperature. {The last line gives the familiar result in terms of electron
pressure relevant for hot electrons, for example, in the clusters of galaxies.} The Compton parameter $\yg$  is a measure of the degree of
comptonization including the effect of all three relevant processes: recoil, induced recoil and
Doppler broadening (the latter determined by $\yg\Te/T$). The distortion
amplitude $\y$ on the other hand is a measure of the
amount of heating (or cooling). {As an example, we show in
  Fig. \ref{Yrifig} the expected  {post-recombination rate of thermal $\y$-distortion injection
  into the CMB.} Before reionization, we get negative $\y$-type distortions due
  to comptonization with colder electrons which cool much faster with the
  expansion of the Universe \citep{cs2011} and can be identified as
  Bose-Einstein condensation \citep{ksc2012}. Reionization, in addition to
  increasing ionization, also heats up the intergalactic medium. We have
  assumed $\Te=10^6/(1+z)^{3.3}~{\rm K}$ at $z<3$ taking into account the
  contributions from the
  warm hot intergalactic medium (WHIM) \citep{co1999,co2006} and $10^4~{\rm K}$ at
  $z>3$. Total 
   $\y$-type distortion from reionization is  $\yg\DTe \sim  (\kB  \Te/\me c^2) \tau_{ri}\sim
10^{-6}\times 0.1=10^{-7}$ while from
   WHIM at smaller redshifts, {it is expected to be $\sim 10^{-6}$
   \citep{ns2001,ks2002,H2009,sb2010}.} {These $\y$-type distortions should be detectable by PIXIE
\citep{pixie} while the proposed experiment Cosmic Origins Explorer (COrE)
\cite{core} with similar resolution but significantly higher sensitivity than PLANCK
  spacecraft \cite{planck} and ground based experiments ACTPol
\citep{actpol} and SPTPol \citep{sptpol} should   also be
able to detect $\y$-fluctuations in the WHIM.}    We have also assumed that
  reionization starts at $z\sim 20$ and ends at $z\sim 8$. Uncertainties in the
  details of reionization and temperature evolution of the intergalactic
  medium make it impossible, at least at present, to disentangle the pre-recombination and
  post-recombination contributions to the $\y$-type distortions. There will
  also be { contributions from the second order Doppler effect
    from baryon peculiar velocities \citep{hss1994b,cs2004}},
not shown in the figure, before during and after reionization. These
contributions during and after recombination are calculated in \citep{cks2012}.} 

With the above definitions, we can now use the standard form of Kompaneets
equation \citep{k1956} (taking care to distinguish between the actual electron/effective
photon  temperature $T_e$ and
the blackbody temperature $T$ used to define the variable $x$):
\begin{align}
\frac{\partial n}{\partial \yg}=\frac{1}{x^2}\frac{\partial }{\partial
  x}x^4\left(n+n^2 +\frac{T_e}{T}\frac{\partial n}{\partial x}\right).\label{komp}
\end{align}
The first term in the parenthesis describes the downward scattering of photons
due to electron recoil, second term is for induced scattering while the
last term describes diffusion of photons in energy due to Doppler effect
and thus depends on the electron temperature.
At high frequencies  $n\ll 1$ and the induced scattering term can be neglected. If a source
of energy raises the electron temperature such that $T_e\gg T$ for a very short time, $\yg\ll 1$, an
analytical solution, for the initial condition $n(x,\yg=0)=\npl(T)$
(blackbody spectrum at temperature T), 
 of the Kompaneets equation can be obtained by approximating the recoil terms
 $n+n^2$ with the initial blackbody spectrum $\npl+\npl^2$.  
The fact that the { Planck (and in general Bose-Einstein)
spectrum is an equilibrium solution of the Kompaneets equation
 (\cite{k1956},\cite{wey1965},\cite{is1975b}) gives $n+n^2\approx -\id n/\id x$.} The
 resulting equation can be transformed into heat diffusion equation , the
 analytic solution of which is the well known $\y$-type distortion \citep{zs1969}{, Eq. \eqref{linsz}.}  We  derive these results in
 Appendix \ref{appy} in a
 way which clearly illustrates the regime of validity of the solution. 
\end{section}

\begin{section}{Evolution of $\y$-type distortion}
We will now consider the problem of comptonization of an initial $\y$-type
distortion created by an energy source which turned on for a very short
time, i.e., instantaneous energy injection.   This problem will 
illustrate the main physics we want to investigate. We will explore the more
realistic cases of continuous energy injection in the next sections.  However, it  is  possible
that such short lived sources may actually exist. {In fact, one such
example  can be found  in standard cosmology.} The decay of primordial $^7Be$ to
$^7Li$ lasts for a very short time at  $z\sim 30000$ \citep{ks2011} and
 gives rise to exactly the type of distortions we calculate in this
section (although energy released in Be decay is too small to be
of observational interest). The spectrum we get from $^7Be$ is the $\yg=0.04$
spectrum and is given approximately by  Eq. \eqref{aneq}  with
$\y\sim 10^{-16}$.

The
$\y$-type spectrum should evolve towards the Bose-Einstein equilibrium solution
with time.
The  equilibrium  electron temperature or effective photon temperature
$T_e$  is given by \citep{zl1970,ls1971}
\begin{align}
\frac{T_e}{T}=\frac{\int(n+n^2)x^4dx}{4\int n x^3 dx}\label{te}
\end{align}
 The effective temperature for the
linear  $\y$-type spectrum for energy injection $\Delta E/E$  is given by \citep{is1975b}
$T_e=T(1+5.4 \y)=T(1+1.35 \Delta E/E)$. The exact temperature
will be slightly higher 
if $\y$ is not small. {For a Bose-Einstein spectrum  the effective
temperature is $T_e\approx T(1+0.456\mu)\approx T(1+0.64\Delta
E/E)$.} This temperature is established very fast compared to any other
relevant process  with a characteristic time of
$\sim 1 \rm{s}$ at $z=10^5$, $\sim 10$ orders of magnitude faster than the
expansion rate of the Universe at that time.
The electrons will thus always be maintained at the effective
photon temperature given by Eq. \ref{te}.  As the $\y$-type spectrum evolves towards the Bose-Einstein spectrum, 
the electron temperature should decrease. 

\subsection{Numerical solution in the intermediate era, $1.5\times
  10^4\lesssim z\lesssim 2\times 10^5$}
\begin{figure}
\resizebox{\hsize}{!}{\includegraphics{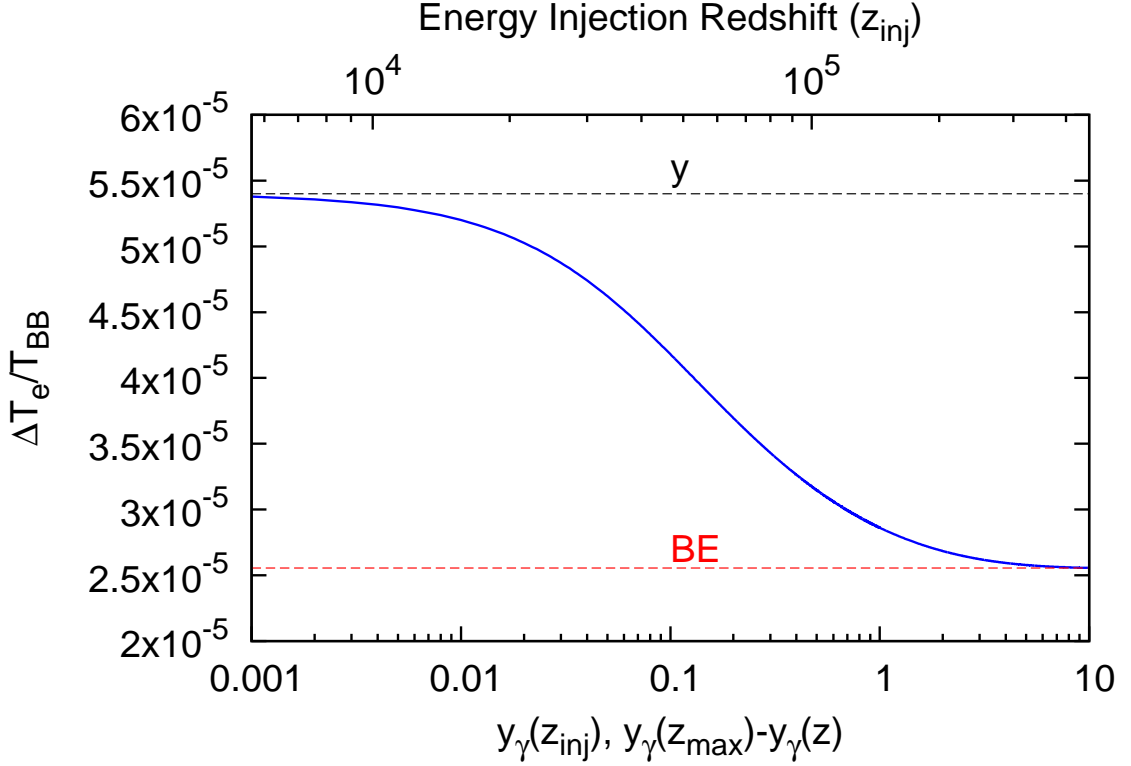}}
\caption{\label{tfig}Evolution of electron temperature for initial
  spectrum with $\y=10^{-5}$. Initial fractional
  difference in temperature $\Delta T_e/T\equiv
  (T_e-T)/T=5.4\times 10^{-5}$. The
  final temperature for the Bose-Einstein spectrum is also marked at
  $\Delta T_e/T=2.56\times 10^{-5}$. By $\yg=0.01$ the temperature is
  significantly different from the initial temperature and by $\yg=2$ it is
  very close to the Bose-Einstein spectrum temperature. {This and subsequent
plots in this section can  be interpreted as snapshots in the evolution of
initial $\y$-spectrum starting from initial energy injection redshift
$\zmax$ to redshift $z$, so that the x-axes is
$\yg=\yg(\zmax)-\yg(z)$. Alternatively, it can be interpreted as the final spectrum
today resulting from the energy injection at redshift $z_{\rm inj}$, so
that $\yg=\yg(z_{\rm inj})$.}
 }
\end{figure}

{To follow the evolution of the spectral distortions starting with the
$\y$-type distortion, we} must solve Eq. \ref{te} and
Kompaneets equation Eq. \ref{komp} simultaneously. Numerically we proceed
as follows. We take small steps in time $\yg$ using Kompaneets equation with
constant $T_e$ given by  Eq. \ref{te} for the spectrum at the beginning of
the step. We then calculate the final electron temperature using
Eq. \ref{te} and 
iterate, with $T_e$ linearly decreasing between the initial and
final values. We found
that a step size of $\delta \yg=0.001$ at $\yg<1$ and $\delta \yg=0.01$ at $\yg>1$ was sufficiently
accurate.  With our iterative procedure the error in energy conservation is $< 1\%$ at
$\yg<10$. Fig.  \ref{tfig} show the cooling
of the photon spectrum as the Bose-Einstein distribution is approached. Initially
 the temperature  drops rapidly from the $\y$-type  value of $\Delta T_e/T=5.4\y$ and
is close to the linear Bose-Einstein value of $2.56\y$ at
$\yg=1$. Fig. \ref{tfig} shows that for small distortions, $\Delta
  n/n\ll 1$ and  $|(\Te-T)/T|\ll 1$, the equilibrium Bose-Einstein spectrum
  is reached at $\yg\gg 1$ and the spectrum is very close to the equilibrium
  at $\yg\sim 1$.
  This conclusion does not depend on the amplitude of the distortion, $\y$,
  at all because the processes responsible for comptonization, Doppler broadening and recoil, are defined by the parameters
$\yg\Te/T$ and $\yg$ respectively.
\begin{figure}
\resizebox{\hsize}{!}{\includegraphics{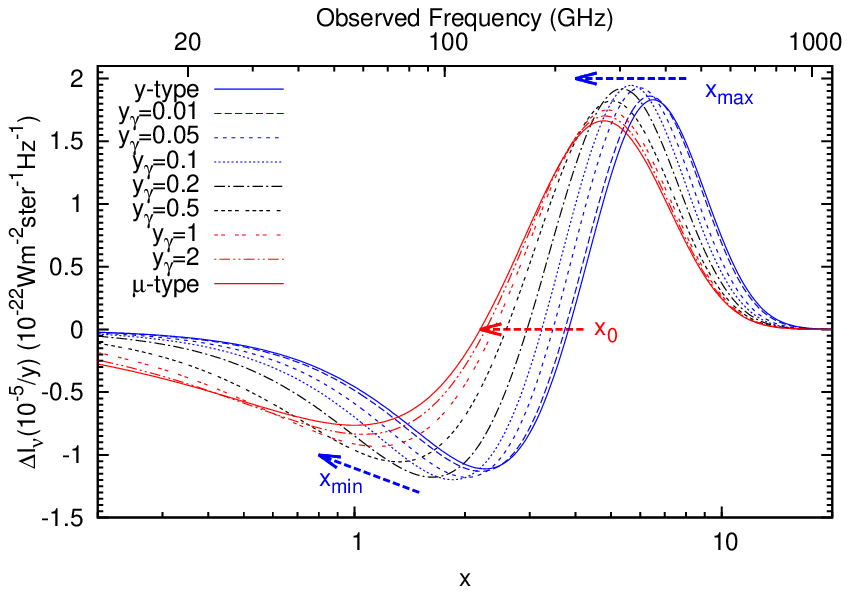}}
\caption{\label{fluxfig}Intermediate-type spectra: Difference in intensity
  from a blackbody at  temperature $T$ is plotted.  Bose-Einstein and
  $\y$-type spectra for
same value of energy injection are also shown. Initial
  spectrum is a pure $\y$-type distortion (Eq. \eqref{linsz} with $\y=10^{-5}$) and is labeled $\yg=0$. The curves in order of increasing
  (non zero) $\yg$
  correspond to energy injection redshift of $z_{inj}=1.56\times
  10^4,3.33\times 10^4,4.67\times 10^4,6.55\times 10^4,1.03\times
  10^5,1.45\times 10^5,2.04\times 10^5$. The 'zero' point, the
frequency 
$x$ where the intensity  equals that of blackbody at temperature
$T$ moves from the $\y$-type distortion value of $3.83$ to Bose-Einstein value of
$2.19$.}
\end{figure}

\begin{figure}
\resizebox{\hsize}{!}{\includegraphics{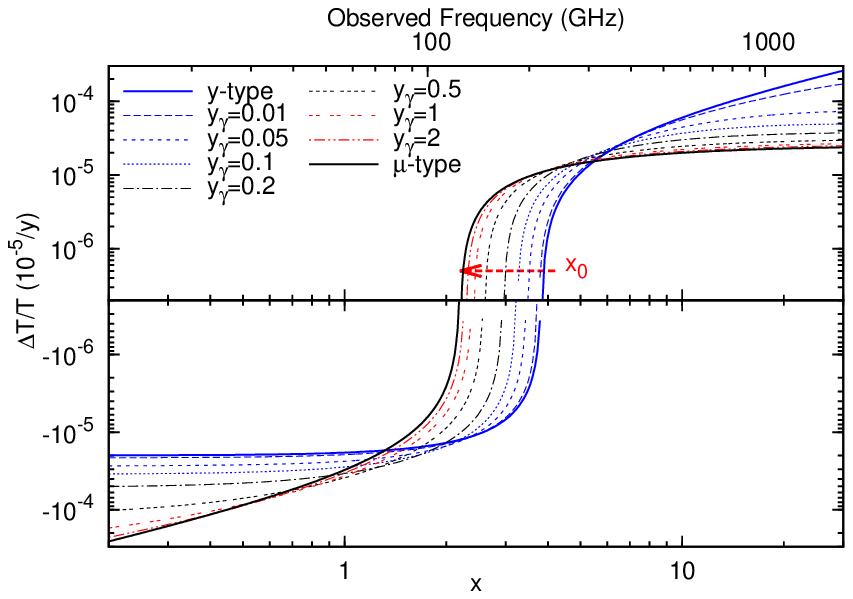}}
\caption{\label{tefffig}Intermediate-type spectra: Fractional difference $\frac{\Delta T}{T}\approx \frac{1-e^{-x}}{x}
  \frac{\Delta n}{\npl}$ in effective temperature (Eq. \eqref{teffeq}) is plotted. Bose-Einstein spectrum for
same value of energy injection is also shown.   Initial
  spectrum is a pure $\y$-type distortion (Eq. \eqref{linsz} with $\y=10^{-5}$) and is labeled $\yg=0$. The curves in order of increasing
  (non zero) $\yg$
  correspond to energy injection redshift of $z_{inj}=1.56\times
  10^4,3.33\times 10^4,4.67\times 10^4,6.55\times 10^4,1.03\times
  10^5,1.45\times 10^5,2.04\times 10^5$. The 'zero' point, maxima and
  minima of the frequency move towards smaller frequencies as
  comptonization progresses. }
\end{figure}

Difference in
observed intensity from a blackbody is shown in Fig. \ref{fluxfig} and 
Fig. \ref{tefffig} 
 shows the fractional difference in the effective temperature  with
respect to $T$ as the photon distribution moves from
the $\y$-type spectrum towards the Bose-Einstein spectrum.   An interesting feature is
that the zero point, defined as $x_0$ such that
$n(x_0)=\npl(x_0)$, moves from the $\y$-type distortion value of $x_0=3.83$ to Bose-Einstein
value of $x_0=2.19$.  The maxima and minima of the intensity distortion also move towards
smaller frequencies as comptonization progresses. Zero point $x_0$ is
plotted in Fig. \ref{x0fig}. Frequencies of maxima and minima, $x_{\rm
  min},x_{\rm max}$ are also plotted
in Fig. \ref{x0fig} and the corresponding intensities, $\Delta I_{\rm
  min}, \Delta I_{\rm max}$ in Fig. \ref{fluxminmaxfig}.
The Bose-Einstein spectrum at $x>10$ is in fact established very quickly. By $\yg=0.2$ the spectrum is very close to the
Bose-Einstein spectrum corresponding to the electron temperature $T_e(\yg)$
at $x>10$. At $\yg>0.2$, the spectrum at high $x$ remains Bose-Einstein and
tracks the electron temperature as the effective radiation/electron temperature
decreases. Fitting formulae for $x_0,x_{\rm min},x_{\rm max}$ are given in
Appendix \ref{appb}. 

\begin{figure}
\resizebox{\hsize}{!}{\includegraphics{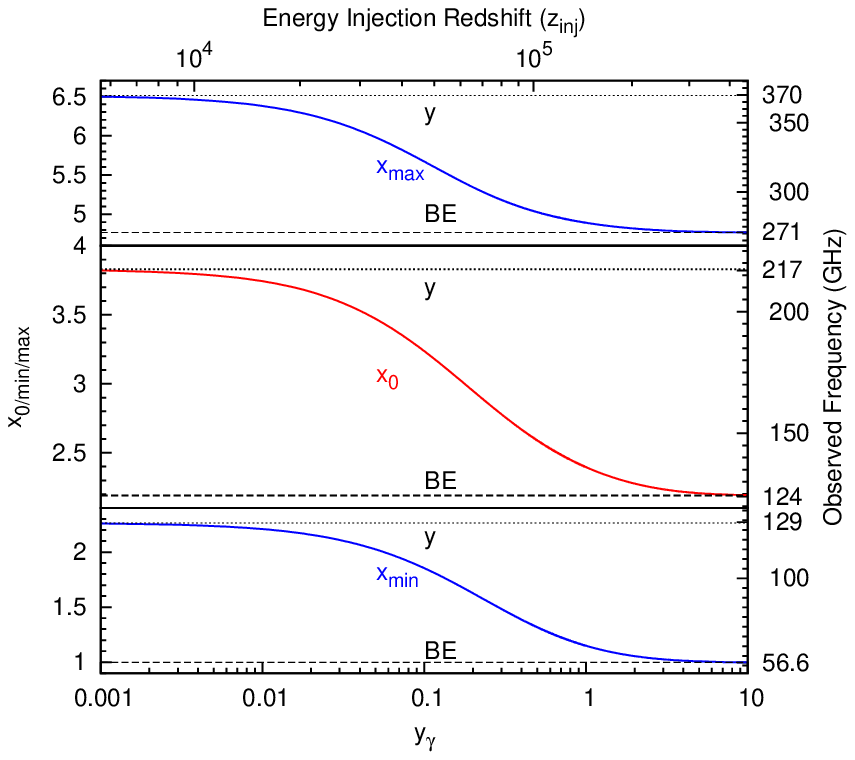}}
\caption{\label{x0fig}Evolution of the zero point $x_0$ defined by
  $n(x_0)=\npl(x_0)$, and frequency of minima and maxima of the intensity
  of distortion, $x_{\rm min}$ and $x_{\rm max}$. $x_0$ can be used to pinpoint the redshift of
  energy injection in case a distortion in CMB spectrum is detected.}
\end{figure}
\begin{figure}
\resizebox{\hsize}{!}{\includegraphics{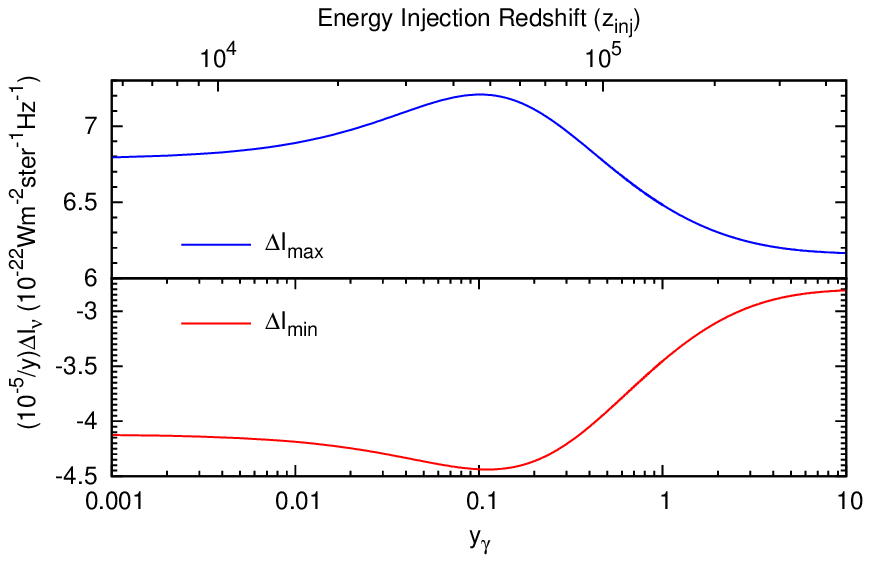}}
\caption{\label{fluxminmaxfig}Minimum and Maximum of the intensity of distortion from the
  reference blackbody, $\Delta I_{\nu}$}
\end{figure}

\end{section}

\subsection{Analytic solution in the weak comptonization limit}\label{appyan}
We can find an analytic solution for the evolution of an initial $\y$-type
distortion  by expanding $n(x,\yg)$ around $\yg=0$, using the Taylor
series expansion, Eq. \eqref{expan}. The initial spectrum is  the $\y$-type spectrum with amplitude $\y$,
$n(x,0)=\npl(x)+\y n_{\y}(x)$, and the initial electron temperature is given by
the equilibrium temperature, Eq. \eqref{te}, $\DTe\approx 5.4 \y$. Substituting the initial spectrum in the
Kompaneets equation gives us the first term in the Taylor series (assuming
$\y\ll 1$ and only keeping terms linear in $\y$), 
\begin{align}
\frac{\partial n}{\partial \yg}(x,0)
&=\frac{1}{x^2}\frac{\partial}{\partial
  x}\left[x^4\left(\frac{\partial n(x,0)}{\partial
      x}\left(\frac{T_{e}}{T}\right)+n(x,0)+n(x,0)^2\right)\right]\nonumber\\
&\approx \DTe n_{\y}(x) +\y f_{\y}(x),\label{linterm}
\end{align}
where $n_{\y}(x)$ is the $\y$-type spectrum defined in Eq. \eqref{linsz}, and $f_{\y}(x)$ is the same function which appeared in Eq. \eqref{szcorr} and is given
in the appendix, Eq. \eqref{fyeq}.
We thus have the solution for the weakly comptonized spectrum,
\begin{align}
n(x,\yg)&\approx \npl(x) + \y\left[n_{\y}(x) +\yg \left(5.4 n_{\y}(x) +
    f_{\y}(x)\right)\right]\nonumber\\
&=\npl(x)+\y n_{\y}(x)\left[1 +\yg \left(5.4  +
    \frac{f_{\y}(x)}{n_{\y}(x)}\right)\right],\label{aneq}
\end{align}
where we have the following simplified expression and large $x$ and small
$x$ limits,
\begin{align}
\frac{f_{\y}(x)}{n_{\y}(x)}&=\frac{8}{x \coth \left(\frac{x}{2}\right)-4}+x
\left[x-\sinh (x)\right] \text{csch}^2\left(\frac{x}{2}\right)+6\nonumber\\
&\xrightarrow{x\gg 1} -2x,\nonumber\\
&\xrightarrow{x\ll 1} 2
\end{align}
The correction $f_{\y}(x)$, of course, similarly to $n_{\y}(x)$, conserves photon number, $\int_0^{\infty} \id x
f_{\y}(x)x^2=0$.

The
 next terms in the Taylor series can be calculated iteratively by taking
 successive $\yg$-derivatives of Kompaneets equation. We give the recursion relations to calculate any higher order
 term in Appendix \ref{appa}.   The  $\yg$-derivatives of
 the electron temperature are also required and are  easily calculated
 using Eq. \eqref{te}. The first two derivatives are 
 given by, $\id \DTe/\id \yg|_{\yg=0} \approx -21.45 \y$, and $\id^2 \DTe/\id
 \yg^2|_{\yg=0} \approx 323.6 \y$. 
 Intermediate distortions for case of continuous
energy release, for example particle decay/annihilation, Silk damping,  are easily obtained from these analytic formulae by linearly
adding (integrating) the spectra for different $\yg$ with appropriate weights.
Analytic solution including first three terms are quite precise ($\sim 1\%$
error) for $\yg\lesssim 0.05$ deteriorating to $\sim 10\%$ errors at
$\yg=0.1$. Numerical and analytic solutions are compared in detail in
Appendix \ref{appa}.

\section{Application: Amplitude, slope and shape of the primordial power
    spectrum on small scales}
{The solutions given in the previous section would be directly applicable if the energy injection occurs over a very short period of time.}
It is more likely, in reality,  that the energy release
happens over an extended period of time, for example, decay of particles or
dissipation of sound waves.  The final spectrum for continuous energy
injection would be a superposition
of spectra for all values of $\yg$ parameter, with appropriate weights
decided by the dependence of rate of energy injection on redshift, and we must calculate the
spectrum for each model of energy release numerically. The shape of the
power spectrum and the value of $x_0$ will still contain information about
the energy release as a function of time, in addition to the total amount
of energy released. This is in contrast to the pure $\mu$-type (or
$\y$-type) distortion which only contains information about the total energy
injected.

\begin{figure}
\resizebox{\hsize}{!}{\includegraphics{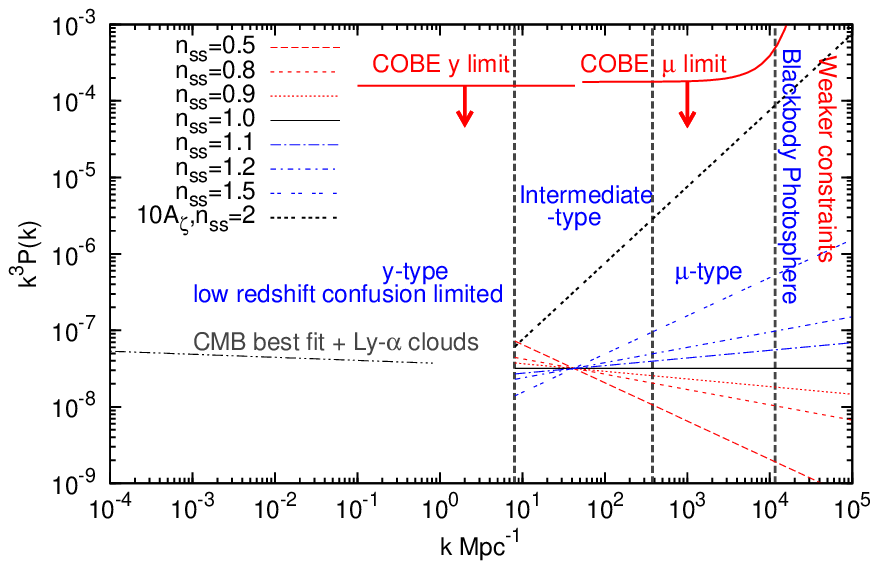}}
\caption{\label{powerfig}The power spectra, Eq. \eqref{powereq},  with different indices $\nSS$
  on small scales are shown, for constant amplitude. Also shown for reference is the best-fit WMAP power
  spectrum on large scales   with $\nS=0.96$ \citep{wmap7,spt,act2012} with
  Ly-$\alpha$ forest extending the constraints to smaller scales \citep{ly2006,ssm2006}. The small scale limits on
  power from COBE/FIRAS measurements of CMB spectral distortions are also
  shown. There is considerable freedom in varying the amplitude and the
  spectral index of the
  power spectrum within COBE/FIRAS limits. We show this by the curve with $10$
  times the amplitude and  extreme value for the spectral index
  $\nSS=2$. {PIXIE \citep{pixie} is expected to improve COBE/FIRAS constraints by a factor of $\sim
  2.5\times 10^3$.}
}
\end{figure}

\begin{figure}
\resizebox{\hsize}{!}{\includegraphics{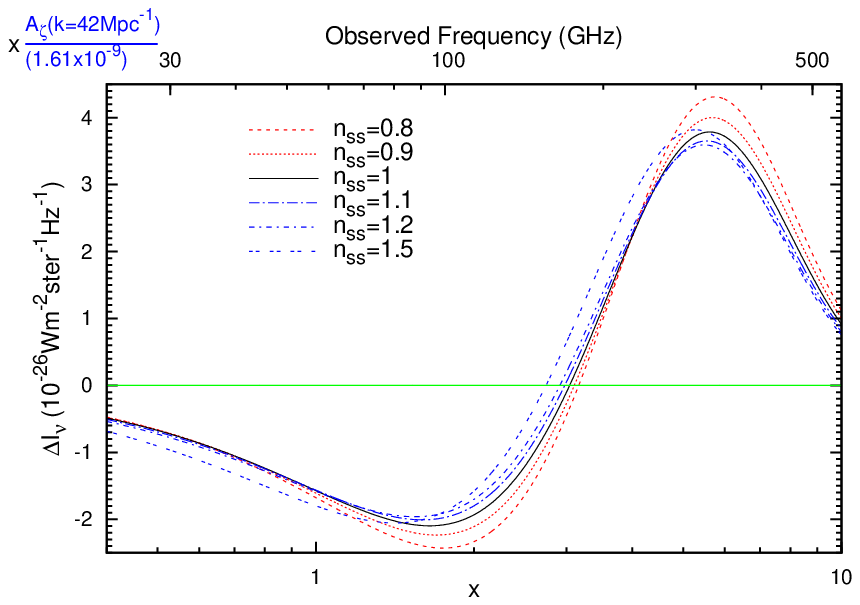}}
\caption{\label{x0nsfig}Distortion spectrum created by dissipation of sound
waves in the redshift range $1.56\times 10^4\le z\le 2.04\times 10^5$
corresponding to the $\yg$ parameter $0.01\le \yg\le 2$. Difference in
intensity from the reference blackbody spectrum  with the same number
density  of photons  is plotted. Different plots are for
different values of small scale spectral index $\nSS$ of the initial power spectrum
normalized so that they all have the same power at the pivot point
$k=42\Mpc^{-1}$ as the WMAP best fit power spectrum with $\nS=0.96$, which is the value of diffusion wavenumber $\kD$ at
$\yg=0.1,z=4.67\times 10^4$. These are the distortions that would be left
over after pure $\yg$ and $\mu$ type distortions are subtracted and probe
\emph{shape} as well as the amplitude of the
small scale power spectrum in the range $8\lesssim (k\sim \kD) \lesssim 378 \Mpc^{-1}$.}
\end{figure}

One of the most important sources of heating in standard cosmology is the
dissipation of sound waves in the early Universe because of the shear viscosity
(and at late times also due to thermal conduction) on small scales. The
dissipation of sound waves resulting from thermal conduction was first
calculated by Silk \citep{silk}, Peebles and Yu \cite{Peebles1970} included shear viscosity
and Kaiser \cite{kaiser} included the effect of photon polarization. The spectral
distortions arising from the dissipation of sound waves have also been
studied by many authors in the past using approximate estimates
\citep{sz1970b,daly1991,hss94} and a precise calculation was done recently
by \citep{cks2012}. {Primordial perturbations excite standing sound waves in
the early Universe on scales smaller than the sound horizon \citep{lifshitz,sakharov66,sz1970c}.} Diffusion of
photons from different phases of the waves, of wavelength of the order
of diffusion  length, gives rise to a local quadrupole which is isotropized
by Thomson scattering (shear viscosity). The effect on the photon spectrum
is just the averaging of the blackbodies of different temperature and gives
rise to  $\y$-type distortions \citep{zis1972}. The $\y$-type distortions can
then comptonize, fully or partially, giving rise to a $\mu$-type distortion
or an intermediate-type spectrum \citep{sz1970b,hss94}.  We should also mention that the adiabatic cooling of baryons due
to the expansion of the Universe gives spectral distortions of an amplitude
opposite to those given by  heating \citep{cs2011}. We include this
cooling of baryons (or equivalently small difference of electron temperature from the
effective photon temperature given by Eq. \eqref{te}) in our calculations; this is, however, a very small correction to the amount of
heating considered below. Also, the low frequency spectrum is affected by
bremsstrahlung emission/absorption after recombination \cite{cs2011}. In the frequency
range of interest to us, $\nu\gtrsim 30~ {\rm GHz}, x\gtrsim 0.5$ and for
distortions of interest, with amplitude $\Delta E/E\gtrsim 10^{-9}$,
low redshift bremsstrahlung (and double Compton scattering) can be neglected.

The precise total spectrum resulting from sound wave dissipation was calculated
recently by
\citep{cks2012} including contributions from the $\mu$-type era,
intermediate era and the $\y$-type era. Here we consider the possibility
that the pure $\mu$-type distortions created at $\yg>\yg_{\rm max}=2 (z\gtrsim 2\times 10^5)$ and pure $\y$-type
distortions created at $\yg< \yg_{\rm min}=0.01 (z\lesssim 1.6\times 10^4)$ can be
subtracted with high precision (see Fig. \ref{fluxfig}). The exact upper/lower limits (and the
resulting intermediate spectrum) will of course
depend on the ability of the experiment to distinguish a pure $\mu$-type ($\y$-type)
from a $\yg_{\rm max} (\yg_{\rm min})$ intermediate-type spectrum. The heating rate
 due to an
initial power spectrum with constant scalar index on small-scales $\nSS$ is given by \citep{cks2012,ksc2012b}
\begin{align}
\frac{\id \SQ}{\id z}=\frac{3.25 A_{\zeta}}{k_0^{\nSS
    -1}}\frac{\id(1/\kD ^2)}{\id z}2^{-(3+\nSS )/2}\kD ^{\nSS
  +1}\Gamma\left(\frac{\nSS+1}{2}\right),\label{energyEq}
\end{align}
where $\SQ=E/\rho_{\gamma}$, $E$ is the total energy in photons and
$\rho_{\gamma}=\aR \TCMB^4(1+z)^4$ is
the reference photon energy density, $\aR$ is the radiation constant,
$\TCMB=2.725~{\rm K}$ is the CMB temperature today, $\kD$ is the damping wavenumber given by \citep{kaiser,weinberg}
\begin{align}
\frac{1}{\kD ^2}&=\int_z^{\infty}\id z
\frac{c(1+z)}{6H(1+R)\Ne \sigT}\left(\frac{R^2}{1+R}+\frac{16}{15}\right)
\end{align}
where $R\equiv3\rho_b/4\rho_{\gamma}$, $\rho_b$ is the baryon energy
density. We have defined the power spectrum of initial curvature
perturbation in comoving gauge $\zeta$ as
\begin{align}
P_{\zeta}=A_{\zeta}\frac{2\pi^2}{k^3}\left(\frac{k}{k_0}\right)^{\nSS-1} \label{powereq}.
\end{align}
An important point to note here is that for $\nSS=1$, the energy released
between redshifts $z_1$ and $z_2$ is proportional to $\ln[(1+z_1)/(1+z_2)]$
\citep{ksc2012}
and it is easily seen that the total energy released at $z\gtrsim 1000$ is
divided approximately equally between $\y$, $\mu$ and intermediate-type
distortions. 
For smaller spectral index $\nSS<1$, intermediate-type distortions get bigger and bigger
share of the total energy released with $\y$-type distortions comparatively
enhanced and $\mu$-type distortions comparatively suppressed. The exact
opposite, of course, happens for larger spectral indices $\nSS>1$.

Since we want to compare the shape of spectral distortions for different
power spectra with similar total energy input,
we choose the pivot point $k_0=\kD(\yg=0.1,z=4.67\times
10^4)=42\Mpc^{-1}$ corresponding to the approximate (geometric) middle
of our redshift range for intermediate-type  spectral distortions. 
We
choose the amplitude $A_{\zeta}=1.61\times 10^{-9}$ to match the power at
$k=42\Mpc^{-1}$ with the WMAP best fit power spectrum with $\nS=0.96$
\citep{wmap7}. The spectral index, $\nSS$, on small scales,  $k\gtrsim 8
\Mpc^{-1}$ can be very different
from the spectral index on large scales, $\nS$, measured by WMAP, for example, in
the case of a running spectrum. The small-scale  power spectra, with different indices
$\nSS$, are shown in Fig. \ref{powerfig}. CMB constrains the primordial
power spectrum only on large scales \citep{wmap7,spt,act2012}, $k\lesssim 0.2 ~{\rm
  Mpc^{-1}}$.  This range can be extended to $\sim 1 ~{\rm Mpc^{-1}}$ using
Ly-$\alpha$ forest \citep{ly2006} and is consistent with the WMAP best fit
power spectrum parameters \citep{ssm2006}.
The small scales are best constrained by COBE/FIRAS data through limits on $\y$ and
$\mu$-type distortions \citep{cobe}. We have shown these constraints, assuming
$n_{ss}=1$ and using the fitting formulae in \citep{cks2012} to calculate the
$\y$ amd $\mu$ type distortions. These constraints are more than 3 orders of
magnitude higher than a simple extrapolation of the WMAP best fit power
spectrum to small scales. Thus, there is considerable freedom, from an
observational viewpoint,  for  the power
spectrum on the small scales to be quite different from the extrapolation of
the WMAP power spectrum to these scales. The constraints get considerable
weaker behind the blackbody surface at small scales,  $k\gtrsim \kD(z=2\times
10^6)=1.15\times 10^4~{\rm Mpc^{-1}}$, because of the suppression of the
$\mu$ distortion by blackbody visibility function \citep{sz1970}. The
visibility  function is  dominated
by the double Compton process in a low baryon density Universe such as ours
\citep{dd1982} and is given by $\approx \exp\left[-(z/2\times
  10^6)^{5/2}\right]$.\footnote{Analytic solution including comptonization
  and both
  bremsstrahlung and double Compton processes with percent level accuracy
  is given in \citep{ks2012}.}. We sketch this weakening of the
constraints  by multiplying the COBE/FIRAS constraint by the inverse of the
visibility function $\exp\left[(k/1.15\times 10^4 ~{\rm Mpc^{-1}})^{5/3}\right]$, using
$\kD\propto (1+z)^{3/2}$. The $\mu$-type distortion, of course, only provide
integrated constraints on the total energy injected in the $\mu$-distortion and its
separation  into individual contributions  from different epochs in
not possible in practice.

To calculate the spectrum arising from a continuous source of heating, such
as dissipation of sound waves, we should solve the Kompaneets equation
Eq. \eqref{komp} with a source term on the right hand side given by 
\begin{align}
\left.\frac{\id n}{\id \yg}\right|_{\rm source}=\frac{1}{4}\frac{\id \SQ}{\id
  z}\frac{\id z}{\id \yg} n_{\y},
\end{align}
where $ n_{\y}$ is the $\y$-type distortion given by Eq. \eqref{linsz}
which is created initially and
the factor of $1/4$ comes from the relation between the energy injected and the
amplitude $\y$ of the $\y$-type distortion. 
We show the intermediate-type spectrum resulting from the dissipation of sound
waves for several different values of the small scale spectral index
$0.5<\nSS<1.5$  in Figs. \ref{x0nsfig} (intensity). The
amplitude of the distortion at low and high frequencies is similar since
the total energy released is similar in all cases. The shape of the
spectrum for different values of spectral index is also similar in the
Rayleigh-Jeans and Wien tails, as expected for the intermediate type
spectra, Fig. \ref{fluxfig}. 
But the spectra are
very different and easily distinguishable near the zero crossing, which
is at   $x_0=3.04,\nu=173 ~{\rm GHz}$ for $\nSS=1$. For $\nSS>1$ there is more
power on smaller scales which dissipate earlier moving the $x_0$ towards
lower values (or towards $\mu$-type value of $x_0=2.19,\nu=124 ~{\rm GHz}$)
and for $\nSS=1.5$, $x_0=2.74,\nu=156~{\rm GHz}$. On the other side, for
$\nSS<1$, 
the spectrum moves towards the $\y$-type value of $x_0=3.83,\nu=217~{\rm GHz}$, and
for $\nSS=0.5$ the zero crossing is at $x_0=3.32,\nu=189~{\rm GHz}$. There
is of course more information in the full spectrum than just the zero
crossing and a sensitive experiment should be able to use the full spectrum
to tightly constrain more complicated shapes of the small scale primordial
spectrum than the simple two parameter (amplitude and spectral index) model
considered here. The zero crossing, and frequencies of minimum and maximum
flux difference with respect to the reference blackbody can be fitted by
the following simple formula as a function of $\nSS$ for $0.5\lesssim
\nSS\lesssim 1.5$ at better than $1\%$ accuracy,
\begin{align}
x_{0/{\rm min/max}}(\nSS)=a_0+a_1 \nSS,
\end{align}
where $a_0=3.61,a_1=-0.588$ for $x_0$, $a_0=2.11,a_1=-0.466$ for $x_{\rm
  min}$ and $a_0=6.24,a_1=-0.63$ for $x_{\rm max}$ 
\section{Application: annihilation and decay of particles}
There are many sources of heating possible in the early Universe from
particle physics beyond the standard model, as discussed in the
introduction. Different sources of energy injection may have different
dependence on redshift and will  give rise to different shapes of
intermediate-type spectrum. We thus have a way of distinguishing between
different types of energy injection mechanisms. We will illustrate this by
considering annihilation of weakly interacting massive particles (WIMP dark
matter), which has a power law dependence on redshift/time, and decay of
unstable particles having an exponential dependence on time.

\begin{figure}
\resizebox{\hsize}{!}{\includegraphics{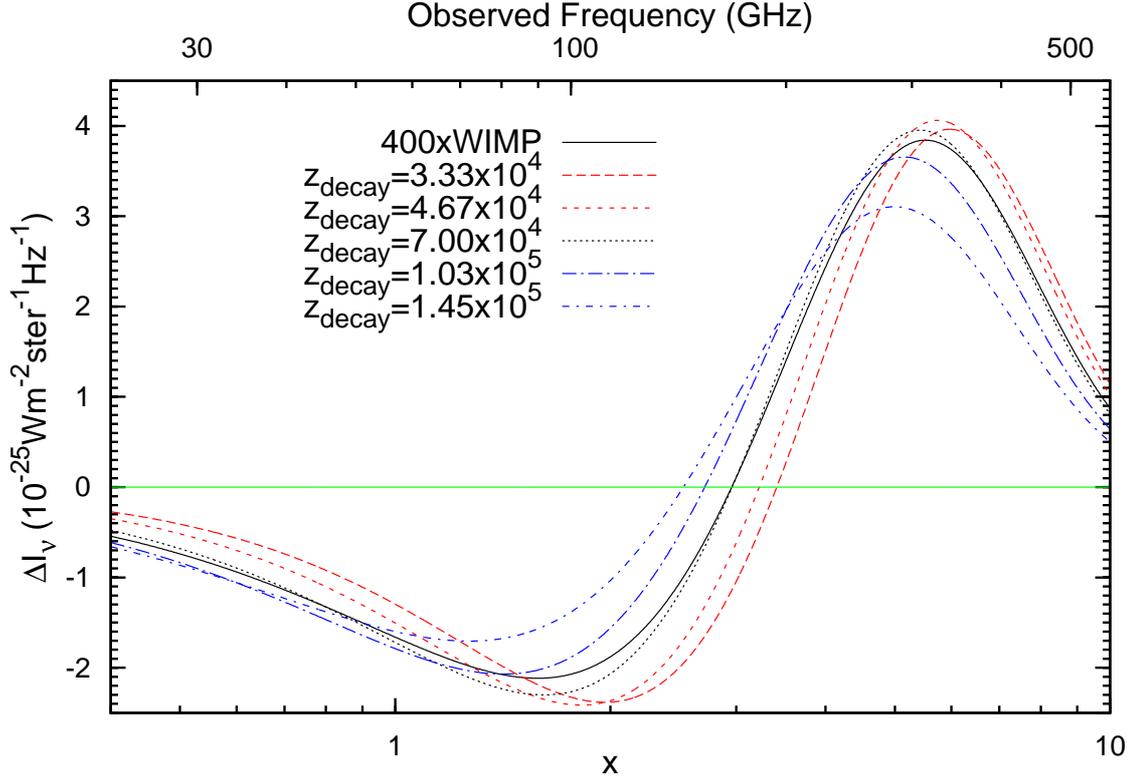}}
\caption{\label{x0particlefig}Distortion spectrum created by annihilation of WIMP
  dark matter (multiplied by 400 to bring its amplitude  to $\sim 10^{-7}$) and decay of unstable particles with different decay times
  $(z_X,\tau_X)=(3.33\times 10^4,661~{\rm  years}),(4.67\times
  10^4,339~{\rm  years}),(7.00\times 10^4,152~{\rm  years}),(1.03\times
  10^5,70.5~{\rm  years}),(1.45\times 10^5,35.7~{\rm  years})$; the
  curves are labeled by $z_X$.  The shape of the WIMP annihilation spectral
distortion is close to the $\nSS=1$ spectrum in
Fig. \ref{x0nsfig}. Difference in intensity from the reference blackbody
with same number density of photons is plotted.}
\end{figure}

For the thermally produced WIMP dark matter consisting of self-annihilating
Majorana particles,\footnote{For Dirac particles the energy release is
  smaller by a factor of 2.} the energy release due to
annihilation is given by,
\begin{align}
\frac{\id \SQ}{\id z}&=-f_{\gamma}\frac{m_{\rm dm}c^2n_{\rm
  dm}^2<\sigma v> }{\rho_{\gamma}(1+z)H}\nonumber\\
&\approx -f_{\gamma}\frac{6.9\times
  10^{-10}(1+z)^{-1}}{\sqrt{1+(1+z_{eq})/(1+z)}}\left(\frac{10~{\rm GeV}}{m_{\rm dm}}\right), 
\end{align}
where we have assumed velocity averaged  cross section $<\sigma v>\approx
3\times 10^{-27}/(\Omega_{\rm dm}h_0^2)~{\rm cm}^{-3}s^{-1}$ \citep{dm},
$\Omega_{\rm dm}$ is the dark matter density as a function of critical
density today, $H$ is the Hubble parameter, $h_0=H_0/100=0.702$, $f_{\gamma}$ is
the fraction of energy going into heating the plasma, $m_{\rm dm}$ is the
mass of dark matter particle, $n_{\rm dm}$ is
the dark matter number density and $z_{\rm eq}\approx 3234$ is the redshift of matter
radiation equality. At $z\gg z_{\rm eq}$ we have $\id \SQ / \id z
\propto 1+z$. For the dissipation of
sound waves in Eq. \eqref{energyEq} under same approximation we have  $\id \SQ / \id z
\propto (1+z)^{(3\nSS-5)/2}$. For $\nSS=1$ dark matter annihilation and
sound wave dissipation have the same redshift dependence and  we expect
 the shape of intermediate-type spectrum for these two cases. More
 precisely, for dark matter annihilation, approximately $30\%$ of the energy
 released at $z\gtrsim 500$ appears as $\mu$-type distortion, $37\%$ as intermediate-type
 distortion and the rest of the energy goes to the $\yg-$type distortion.

For decay of a particle of mass $m_X$, initial comoving number density
$n_{X0}$, and life time $\tau_X$ with $f_{\gamma}$
fraction of energy going into heating of the plasma, we have
\begin{align}
\frac{{\id \SQ}}{\id z}=-f_{\gamma}\frac{n_{X0}m_Xc^2e^{-t/\tau_X}}{\aR
  \TCMB^4 H (1+z)^2\tau_X}
\end{align}

To get total distortion of $\sim 10^{-7}$, we choose $f_{\gamma}n_{X0}m_Xc^2=10^{-7}\aR
  \TCMB^4 (1+z_X)$, where $z_X$ is the decay redshift corresponding to the
  lifetime $\tau_X$, and during radiation domination we have $\tau_X\approx
  1/(2(1+z_X)^2H_0\Or^{1/2})$. The division of released energy
    into $\y$, $\mu$ and intermediate-type distortions varies quite
    dramatically  with the lifetime of the particle and illustrates nicely
    how using the intermediate-distortions along with $\mu$ and $\y$-type
    distortions can help remove degeneracies
    associated with different energy release mechanisms. For $z_X=1.45\times 10^5$, $21\%$ if
  released energy appears as $\mu$-type distortion and $79\%$ as
  intermediate type distortions. For $z_X=4.67\times 10^4,7\times 10^4$, almost all of the
  energy, $\sim 99\%, 97\%$ respectively, 
   goes to intermediate-type distortions with the most of the rest in
   $\mu$-type distortions. For $z_X=1.5\times 10^4$ the division is $42:58$
   between intermediate and $\y$-type distortions respectively and
   $\mu$-type distortions get a negligible share.

The intermediate-type spectral distortion for WIMP annihilation and
  decay of an unstable particles with different lifetimes $z_X$ is shown in
  Fig. \ref{x0particlefig} (intensity). The WIMP spectral distortion is multiplied by $400$ to bring
  it to the same level as the decaying particle signal. WIMP annihilation
  spectrum is similar to $\nSS=1$ spectrum in Fig. \ref{x0nsfig} as
  expected from their similar redshift dependence. Also, the WIMP annihilation
    (power law dependence of energy injection on redshift) and the spectra for decaying
    particles (exponential dependence on redshift) with different lifetimes
    are  distinguishable from each other. There is
   a small degeneracy for $z_X\approx 7\times 10^4$; the exponential decay
  in this case has same zero crossing as annihilation, and the shapes of
  two curves are very close. The $\mu$-type distortions in the two
  cases are very different, as discussed above, and break this degeneracy. Nevertheless, the intermediate-type spectrum has  the
  possibility of \emph{measuring} the life-time of the decaying particle in
  addition to the total energy injected into the CMB.
The frequencies $x_0,x_{\rm min},x_{\rm max}$ can be fitted by the formula
for dark matter decay at with $z_4\equiv z_{X}/10^4$ for
$10^4\lesssim z_{X}\lesssim 2\times 10^5$ with better than $1\%$ precision,
\begin{align}
x_{0{\rm min/man}}=a_0+a_1\ln(z_4) +a_2\ln^2(z_4)+a_3\ln^3(z_4)+a_4\ln^4(z_4),
\end{align}
where for $x_0$ the fit coefficients are
$a_0=3.67,a_1=-0.0644,a_2=-0.017,a_3=-0.1365,a_4=0.0343$, for $x_{\rm min}$
they are
 $a_0=2.16,a_1=-0.07,a_2=0.0146,a_3=-0.1165,a_4=0.0276$, and for $x_{\rm min}$
we have  $a_0=6.25,a_1=0.0158,a_2=-0.157,a_3=-0.1,a_4=0.0335$. For dark
matter annihilation, the frequencies are $x_0=2.96,x_{\rm min}=1.58,x_{\rm max}=5.52$.
\section{Non-degeneracy among Intermediate-type distortions and a
  mixture of $\y$ and $\mu$-type distortions}
Since the intermediate type distortions lie in-between $\y$ and $\mu$ type
distortions, an important question arises: can an intermediate-type
contribution to the CMB spectral distortion be mistaken for a combination
of $\y$ and $\mu$ type distortions and vice versa?
The answer is no, these three types of distortions are non-degenerate with
each other and can be distinguished. This is clear by looking at
Fig. \ref{fluxfig}. The intermediate distortions ($\Delta T/T$) for $0.01\le \yg \le 2$
are almost constant at both low and high frequencies. The $\y$-type
distortions, on the other hand,  rise  at high frequencies as
$\Delta T/T \propto x$. The magnitude of the $\mu$-type distortions
similarly  increases  at low
frequencies with $\Delta T/T \propto 1/x$. Any mixture of pure $\y$ and
$\mu$ type distortions will thus have much greater slopes than the
intermediate type distortions, and in principle they  can be
separated from each other.

\begin{figure}
\resizebox{\hsize}{!}{\includegraphics{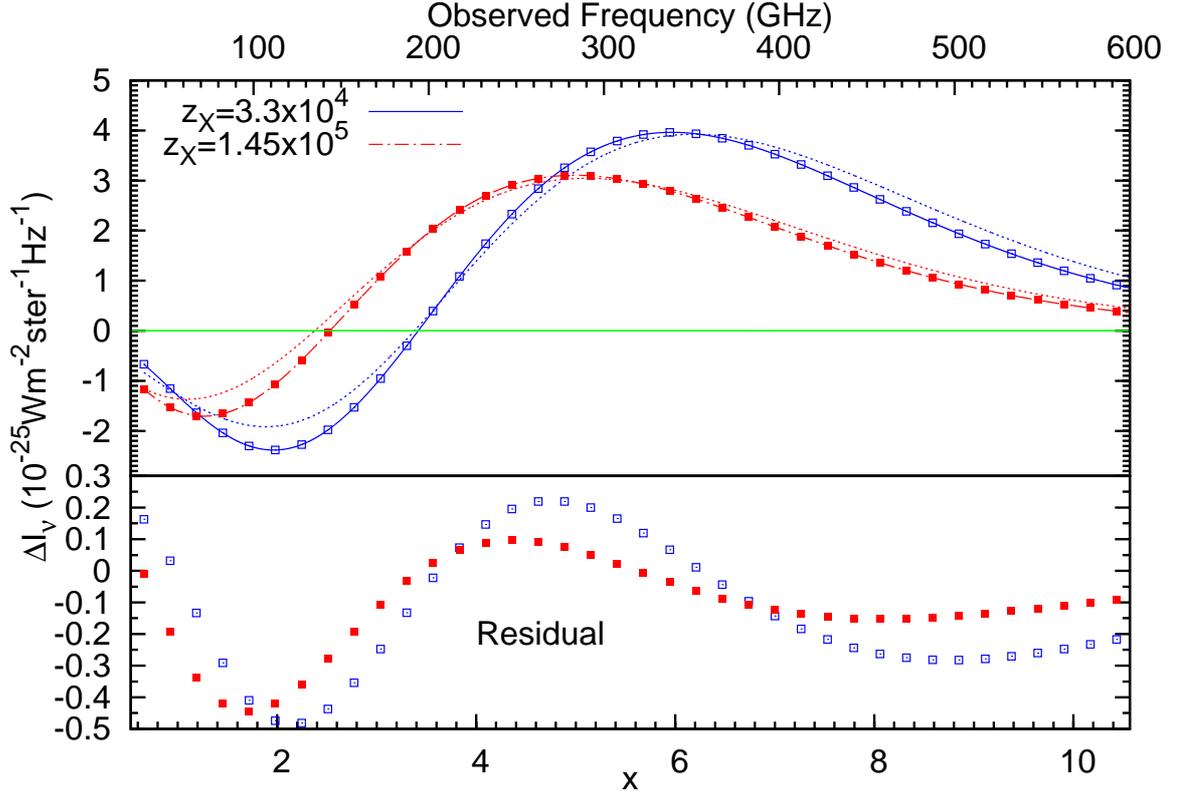}}
\caption{\label{mimicfig}Intermediate-type distortions from particle decay are shown for
  two different decay times, as in Fig. \ref{x0particlefig}. Also shown
  is the least squares fit of  the mixtures of $\y$ and $\mu$ type distortions (dotted curves) which approximate
  these intermediate-type distortions at PIXIE frequencies. }
\end{figure}

We show in Fig. \ref{mimicfig} the same spectrum as in
Fig. \ref{x0particlefig}, for the energy injection due to particle decay for
two different decay times, $z_X=1.45\times 10^5,
   3.33\times 10^4$. Also shown is a combination of $\mu$ and
  $\y$-type distortions, which approximate these intermediate-type spectral
  distortions in the least squares sense.  Thus, a sensitive experiment should be
  able to distinguish between $\mu$, $\y$ and intermediate-type
  distortions when the distortions are detected at high significance. In particular, it should be possible
  to not only avoid the contamination of  the $\mu$ type distortions from
the   intermediate-type distortions but also measure the intermediate-type
  distortions themselves.

\begin{figure}
\resizebox{\hsize}{!}{\includegraphics{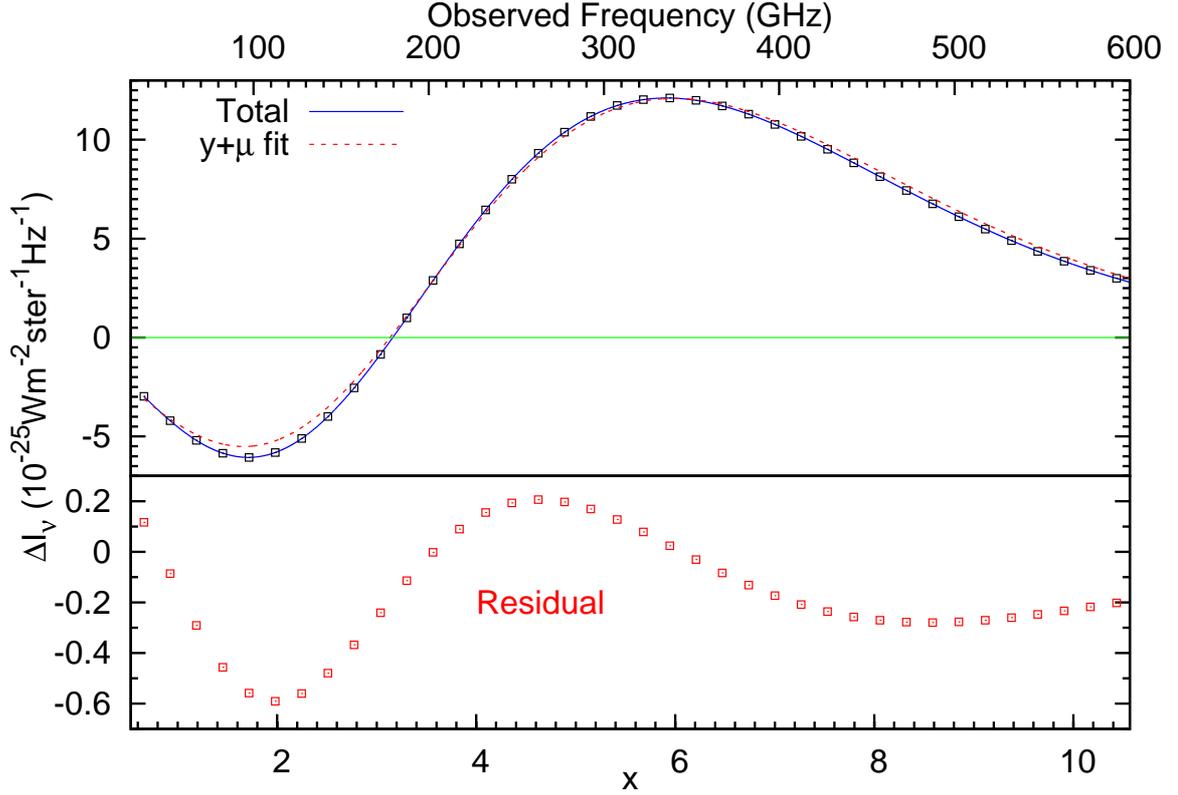}}
\caption{\label{interfig} Total spectrum containing an intermediate-type
  distortion from Silk damping with $\nSS=1$, and amplitude
  $10A_{\zeta}=1.61\times 10^{-8}$, and $\mu$ and $\y$-type distortions with
  $\mu=10^{-7}$ and $\y=3\times 10^{-8}$. All three components
  have approximately the similar amount of energy. Also shown is the best fit
  $\yg+\mu$ spectrum to the data points at PIXIE frequencies. The best fit
  spectrum has $\mu=1.8\times 10^{-7}$ and $\y=4.26\times 10^{-8}$. The
  residual (data-fit) is shown in the bottom panel.}
\end{figure}

The issue of degeneracy and the sensitivity required to detect intermediate
type distortions can be made more precise by asking a slightly different
question: How closely can a  total spectrum containing all $\mu$, $\y$ and
intermediate-type distortions can be fitted by just $\mu$ and $\y$-type
distortions. We show in Fig. \ref{interfig} the total spectrum from Silk
damping with $\nSS=1$, and amplitude $10A_{\zeta}=1.61\times 10^{-8}$. The
total spectrum has a
Bose-Einstein part with $\mu= 10^{-7}$ and $\y$-type part with
$\y=3\times 10^{-8}$. In reality, the $\y$-type part of the spectrum would be
much higher because of the contributions from reionization and later times,
but this does not change our arguments or conclusions. Also shown is the least squares fit using the data
points  at PIXIE frequencies (also shown) to the spectrum with only $\y$
and $\mu$-type distortions. The best fit spectrum has $\y=4.26\times
10^{-8}$ and $\mu=1.8\times 10^{-7}$. The bottom panel shows the
difference between the data points and $\y+\mu$ fit. It is clear from these
plots that the full spectrum cannot be fit exactly with only $\y$ and
$\mu$-type components and the residuals (data-fit) contain give information about the
type of intermediate-distortion present. {In particular the residuals, which
are $\sim 20\%$ of the intermediate-type distortion, are
not affected by the presence of additional $\y$ and $\mu$ components in the spectrum. Thus the detection of the intermediate-type spectrum will be
challenging but possible. 
In addition, since  we do
not know the temperature of the blackbody part of the
spectrum at the required precision a priori, we should  fit for the
temperature of blackbody along with the distortions \cite{cobe}.} 

\section{Observational issues}
Our definition of spectral distortions and $x_0$ is 
convenient to compare and understand the process of comptonization and theoretical spectra of different types.  
Observationally, it is not possible  to calculate the reference blackbody temperature
required for these definitions to desired accuracy. Achieving an accuracy
of $\sim 10^{-8}$ in the number density of photons, and hence the reference
temperature, requires integrating the spectrum between $10^{-4}\lesssim x
\lesssim 25$! This  means that $x_0,x_{\rm min},x_{\rm max}$ as defined by us is not a
direct observable for small distortions. $x_0,x_{\rm min},x_{\rm max}$ can
however be inferred once the full spectrum is measured with sufficient precision
and the
intermediate-type spectral distortion part separated.

The 
  shape of the intermediate-type spectral distortions  contains much more, and
  complementary, information compared to the $\mu/y$-type distortion. High precision
  experiments in the future may thus be able to not only put stringent constraints
  on the energy release in the early Universe using $\mu$ distortions, but
   also  distinguish between different mechanisms of energy
  injection using the intermediate-type spectral distortions.
Detecting the $\mu$, $\y$ and intermediate-type  distortions at the level of
$10^{-7}-10^{-9}$ would require understanding and subtracting foregrounds
at high precision. Simulations for the proposed experiment PIXIE
\citep{pixie} show that such a subtraction may be possible at ${\rm nK}$
level. However, there are still some uncertainties in our understanding of
the  foregrounds as demonstrated by the unexplained excess flux at low
frequencies in the  ARCADE experiment measurements \citep{arcade} and more
work needs to be done to demonstrate the feasibility of the measurements
proposed in the present paper.
 
\begin{section}{Conclusions}
The CMB blackbody spectrum in the $\Lambda CDM$
cosmology is established at $z>2\times 10^6$, when the photon number changing
processes of bremsstrahlung and double Compton scattering are
effective. Compton scattering establishes a Bose-Einstein spectrum
corresponding to the energy and number density  of photons available  while
bremsstrahlung and double Compton scattering help drive the chemical
potential to zero.  If there is energy/photon production due to
non-standard physics at $z<2\times 10^6$, the blackbody spectrum cannot be
restored and  an imprint is left on the CMB
spectrum. Initially a $\y$-type distortion is established which  evolves (comptonizes)
towards a Bose-Einstein spectrum. How close the initial $\y$-type distortion can
get to a Bose-Einstein spectrum depends on the redshift of energy
release. At $z\gtrsim 2\times 10^5$, full comptonization is possible but at
lower redshifts Compton scattering can no longer
establish a 
Bose-Einstein spectrum. At redshifts $1.5\times 10^4 \lesssim z \lesssim
2\times 10^5$, there is partial comptonization and we get a spectrum which
is in-between a $\y$-type spectrum and a Bose-Einstein spectrum. The
detailed shape of this intermediate-type spectrum
 depends on how the rate of energy injection
varies with the redshift.   The detection of a deviation
of the CMB spectrum from a blackbody  thus contains  information about
the amount  as well as the redshift/mechanism of the energy release. 

We have numerically calculated the detailed evolution of the initial $\y$-type 
distortion by solving the Kompaneets equation taking into account the
correct evolution of the electron temperature, which decreases from the
initial $\y$-type distortion value towards a Bose-Einstein value. An
analytic solution valid in the weak comptonization limit, $\yg\lesssim 0.1$,
is given in Eq. \eqref{aneq}. We have
demonstrated  that  the intermediate-type spectral
distortions, resulting from sound wave dissipation in the early Universe
can, in principle, constrain the shape of the small scale power
spectrum. It is also possible to distinguish between different energy
injection mechanisms, which have different dependence on redshift, for
example, WIMP annihilation and Silk damping with different power law dependence and particle decay
with exponential dependence.
{The $\y$, $\mu$ and intermediate-type  
distortions have important differences in  their shapes and it is in principle possible to
distinguish between them. In particular,  a mixture of $\y$ and $\mu$-type
distortions fails to mimic the intermediate-type distortions by $20\%$, and vice versa.
Proposed
experiment PIXIE \citep{pixie}  will detect $\y$ and $\mu$-type distortions at the level of
$\y=10^{-8},\mu=5\times 10^{-8}$ improving the current limits by 3 orders
of magnitude.  } The  additional
$\y$-distortions from low redshifts should not affect an experiment's ability to detect
$\mu$ and intermediate-type distortions. Measurement of intermediate-type
distortions will thus be challenging but possible. 

We make publicly available\footnote{\url{http://www.mpa-garching.mpg.de/~khatri/idistort.html}} high
precision 
intermediate-type distortion templates and a Mathematica code which
superposes these templates, according to user-defined redshift-dependent
energy injection rate, to calculate the $\mu$, $y$ and intermediate-type distortions.
\end{section}
\acknowledgments
We would like to thank Jens Chluba for comments on the manuscript.

\bibliographystyle{JHEP}
\bibliography{cmb_distortions}

\begin{appendix}
\section{Fitting formulae for $x_0,x_{\rm \min},x_{\rm max}$ as a function
  of $\yg$}\label{appb}
$x_0(\yg),x_{\rm min}(\yg),x_{\rm max}(\yg)$ for the intermediate type spectra in
Fig. \ref{fluxfig} are well fitted (for small distortions) with an accuracy better than $1\%$ for
$\yg\lesssim 10$ by the
following formulae:
\begin{align}
x_{\rm 0/min/max}&=a_0+a_1 \yg+a_2\yg^2 +a_3 \yg^3 + a_4 \yg^4+b_1 \sinh^{-1} (b_2 \yg)\nonumber\\
&+c_1 \tanh(c_2 \yg),
\end{align}
where we have  for $x_0$, 
\begin{align}a_0&=3.83,a_1=0.363,a_2=-5.68\times 10^{-2},a_3=5.15\times
10^{-3},a_4=-1.85\times 10^{-4},\nonumber\\
 b_1&=0.496,b_2=-30.8,c_1=0.294,c_2=20.6,
\end{align}
for $x_{\rm min}$
\begin{align}a_0&=2.265,a_1=0.332,a_2=-5.07\times 10^{-2},a_3=4.4\times
10^{-3},a_4=-1.5\times 10^{-4},\nonumber\\
 b_1&=0.439,b_2=-21,c_1=0.24,c_2=14.1,
\end{align}
and for $x_{\rm max}$
\begin{align}a_0&=6.51,a_1=0.543,a_2=-0.11,a_3=1.1\times
10^{-2},a_4=-4\times 10^{-4},\nonumber\\
 b_1&=0.5,b_2=-89,c_1=0.56,c_2=51.
\end{align}

\section{Analytic approximate solutions of Kompaneets equation}
\label{appy}

We expand the photon occupation number ($n(x,\yg)$)  around the initial black
body 
spectrum at temperature $T$, $n(x,0)\equiv \npl(x)\equiv 1/(e^{h\nu/k_BT}-1)=1/(e^{x}-1)$. 
\begin{align}\label{expan}
n(x,\yg)=n(x,0)+\yg\frac{\partial n}{\partial \yg}(x,0) + \frac{\yg^2}{2}\frac{\partial^2 n}{\partial \yg^2}(x,0)+\mathcal{O}(\yg^3),
\end{align}
where $x=h\nu/k_B T$ and $T$ is the initial blackbody  temperature.
Further we can evaluate the corrections using Kompaneets equation.
\begin{align}
\frac{\partial n}{\partial \yg}(x,0)&=\frac{1}{x^2}\frac{\partial}{\partial
  x}\left[x^4\left(\frac{\partial n(x,\yg)}{\partial
      x}\left(\frac{T_{e}}{T}\right)+n(x,\yg)+n(x,\yg)^2\right)\right]_{\yg=0}\nonumber\\
&=\frac{1}{x^2}\frac{\partial}{\partial
  x}\left[x^4\left(\frac{\partial n(x,0)}{\partial
      x}\left(\frac{T_{e}}{T}-1\right)\right)\right]\nonumber\\
&=\DTe\frac{1}{x^2}\frac{\partial}{\partial
  x}\left[x^4\left(\frac{\partial n(x,0)}{\partial
      x}\right)\right]\nonumber\\
&=\DTe\frac{xe^x}{(e^x-1)^2}\left[x\left(\frac{e^x+1}{e^x-1}\right)-4\right]\nonumber\\
&\equiv \DTe n_{\y}(x),
\end{align}
where  $n_{\y}(x)$  is just the linear $\y$-type solution found in \citep{zs1969}. The next term in the Taylor series is
\begin{align}
\frac{\partial^2 n}{\partial \yg^2}(x,0)&=\frac{\partial}{\partial \yg}\frac{1}{x^2}\frac{\partial}{\partial
  x}\left[x^4\left(\frac{\partial n}{\partial
      x}\left(\DTe +1\right)+n+n^2\right)\right]_{\yg=0}\nonumber\\
&=\DTe\frac{1}{x^2}\frac{\partial}{\partial
  x}\left[x^4\left(\frac{\partial n(x,0)}{\partial
      x}\frac{1}{\DTe}\frac{\partial \DTe}{\partial
      \yg}(0)\right.\right.\nonumber\\
&+\left(\DTe\frac{1}{x^2}\frac{\partial }{\partial
      x}x^4\frac{\partial^2 n(x,0)}{\partial
      x^2}\right)\nonumber\\
&\left.\left.+\frac{2}{x}\left(\DTe +1\right)\frac{\partial}{\partial
    x}\left(x^2 \frac{\partial n(x,0)}{\partial
      x}\right)-2x^2\left(\frac{\partial n(x,0)}{\partial
      x}\right)^2\right)\right]\nonumber\\
&\equiv \DTe f_{\y}(x)+\DTe^2 f_2(x)+\frac{\partial \DTe}{\partial
      \yg}(0)n_{\y}(x).\label{szcorr}
\end{align}
 The functions $f_{\y}(x)$ and
$f_2(x)$ are given explicitly in the Appendix \ref{appa}. Substituting the first and second order terms from
  Eqs. \eqref{szcorr} and \eqref{linsz} in Eq. \eqref{expan}, we  have the solution for partial comptonization of an
  initial blackbody spectrum interacting with electrons at temperature
  $\Te(\yg)$, valid for small distortions and correct to second order in $\yg$,
\begin{align}
n(x,\yg)=&\npl+\yg\DTe n_{\y} \nonumber\\
&+ \frac{\yg^2}{2}\left[\DTe f_{\y}(x)+\DTe^2 f_2(x)+\frac{\partial \DTe}{\partial
      \yg}n_{\y}(x)\right]+\mathcal{O}(\yg^3),\label{taylorsol}
\end{align}
where $\DTe$ and its derivative  are evaluated at $\yg=0$.

The last term, $\frac{\partial \DTe}{\partial \yg}$, depends on the physics
responsible for the electron temperature $\Te$.   It is also clear that higher order terms
will have higher order derivatives of $\DTe$ and the above solution is only
valid when these higher order derivatives are negligible compared to the
first derivative. We will ignore this term in the rest of this
section, as it is not relevant to the present discussion. In the early
Universe, however, electron temperature is not constant and changes as
comptonization progresses and we will discuss the evolution of the electron
temperature in detail in the next sections. If the electrons are in
equilibrium with radiation,  $\frac{\partial
  \DTe}{\partial \yg}$ is easily calculated by taking derivative of
Eq. \eqref{te}.  

We can now use the fact that the blackbody spectrum is a steady state
solution of the Kompaneets equation.
\begin{align}
n(x,0)+n(x,0)^2&=-\frac{\partial n(x,0)}{\partial x}\nonumber\\
&=-\frac{\partial n(x,\yg)}{\partial
  x}+\yg\frac{\partial^2 n}{\partial \yg\partial x}(x,0)+
\frac{\yg^2}{2}\frac{\partial^3 n}{\partial \yg^2\partial x}(x,0)\nonumber\\
&+\mathcal{O}(\yg^3)\label{bbexp}
\end{align}
We have used Eq. \ref{expan}  in the last step.
Using again the expansion Eq. \ref{expan} and Eq. \ref{bbexp} in the Kompaneets equation to replace
$n+n^2$ term and , we get, on ignoring terms of order $\yg^3$ and higher,
\begin{align}
\frac{\partial n(x,\yg)}{\partial \yg}&=\frac{1}{x^2}\frac{\partial}{\partial
  x}\left[x^4\left(\frac{\partial n(x,\yg)}{\partial
      x}\left(\frac{T_{e}}{T}-1\right)\right.\right.\nonumber\\
&\left.\left.+\yg\left(\frac{\partial^2 n}{\partial
        \yg\partial x}(x,0)+\frac{\partial n}{\partial
        \yg}(x,0)+2n(x,0)\frac{\partial n}{\partial \yg}(x,0)\right)\right.\right.\nonumber\\
&+\frac{\yg^2}{2}\left(\frac{\partial^3 n}{\partial
        \yg^2\partial x}(x,0)+\frac{\partial^2 n}{\partial
        \yg^2}(x,0)+2n(x,0)\frac{\partial^2 n}{\partial \yg^2}(x,0)\right.\nonumber\\
&\left.\left.\left.+2\left(\frac{\partial n}{\partial \yg}(x,0)\right)^2 \right)\right)\right]+\mathcal{O}(\yg^3)
\end{align}
 We are interested in the behavior of corrections as $\yg$
and $\DTe$ increase. Thus collecting all terms of same order in $\yg$ and
$\DTe$ (and
defining representing $x$-dependence with functions $F_n(x)$ for
brevity)\footnote{We only give  the $x\gg
  limit$ below but the functions $F_n(x)$ just involve the Planck function
  $\npl$ and its  derivatives 
  are easily calculated explicitly if desired.},
 we have
\begin{align}
\frac{\partial n(x,\yg)}{\DTe \partial \yg}\equiv &\frac{1}{x^2}\frac{\partial}{\partial
  x}\left[x^4\frac{\partial n(x,\yg)}{\partial
      x}\right]+ \yg  F_1(x) +\yg^2 F_2(x) +\yg^2\DTe
  F_3(x)\nonumber\\
&+\mathcal{O}(\yg^3)&\label{szeq}\\
\overset{x\gg 1}{\approx} \hspace{8 pt}&\frac{1}{x^2}\frac{\partial}{\partial
  x}\left[x^4\frac{\partial n(x,\yg)}{\partial
      x}\right]- \left(2 \yg x^3    -3\yg^2 x^4+2\yg^2\DTe
    x^5\right)\npl(x)\nonumber\\
&+ \mathcal{O}(\yg^3)&\label{szasym}
\end{align}
The last expression illustrates clearly the regime of validity of the $\y$-type
solution.  The $\y$-type solution is valid in the limit  $\yg\ll 1$ and 
$\yg^2\DTe \ll 1$. Also the solution fails at $x\gg 1$, when the recoil effect 
gives deviations of order unity with respect to the blackbody. The recoil effect would lead to a downward shift in the
frequency of high energy 
photons given by $1/x'-1/x=\yg$ \citep{arons,is72}. Thus the $\y$-type solution is only
valid for $x\ll 1/\yg$. This is also easily seen by comparing the first term on
the right hand side ($\sim x^2 \npl$) with the first two terms in round
brackets, $(2\yg x-3\yg^2x^2)x^2\npl$ in the limit $x\gg 1$.

In particular for $\DTe \ll 1$, as is the case in the
early Universe before recombination, the $\yg\ll 1$ is the stronger condition.
Thus the $\y$-type solution is the correct solution for energy injection at
$z\lesssim 20000$ with the corrections due to higher order terms of the order $\sim \yg \sim 10^{-2}$
(Fig. \ref{yfig}).

With $\y=\int_0^\yg \DTe \id\yg$ and 
 $\y\ll 1$, an approximate linear solution of Eq. \eqref{szeq} follows by evaluating
  the right hand side of Eq. \eqref{szeq} at $\yg=0$,
with  $\y=1/4  (\Delta E/E_r)$ , $\Delta E$ is the energy injected
into the plasma and $E_r$ is the initial radiation density.
\begin{align}
  n(x,\y)-\npl(x)& = \y n_{\y}(x) \nonumber\\
&=\y \frac{xe^x}{\left(e^x-1\right)^2}\left[x\frac{e^x+1}{e^x-1}-4\right]\label{szap}
\end{align}

An important difference between Eq. \eqref{szap} and linear $+$
  second order $\id \DTe/\id \yg$ term
   in
  Eq. \eqref{taylorsol}, although they look identical, is that in Eq.
  \eqref{szap} we have defined $\y$ as an integral over $\DTe$ and there is
  no restriction on the higher order derivatives of $\DTe$, except that the
  integral $\y\ll 1$ and the assumptions under which Kompaneets equation is
  derived are
  valid. We have in
  effect summed over all the terms coming from the Taylor series expansion of
  $\DTe$ in Eq. \eqref{taylorsol}. The solution in Eq. \eqref{szap}, written in terms of $\y$, is
  thus, more generally applicable.
This solution is valid for $\Delta n/\npl \ll 1$, which implies that for large $x$
we have the condition $x^2\y\ll 1$. This condition is clearly satisfied
for the CMB, with the current limit of $\y<10^{-5}$, in the Wien tail for
  $x<100$. 
The well known solution Eq. \eqref{szap} \citep{zs1969} depends only on $\y$
and not on $\yg$. The next correction depends also on $\yg$. This demonstrates
that the broadening of the spectrum and redistribution of photons over the
frequency is defined by $\Te$ but the energy exchange between the plasma and
 the radiation is defined by $\Te-T$.
\section{Corrections to
    $\y$-type distortion from weak comptonization and recursion relations
    for calculating the higher order terms}
\label{appa}
\begin{figure}
\resizebox{\hsize}{!}{\includegraphics{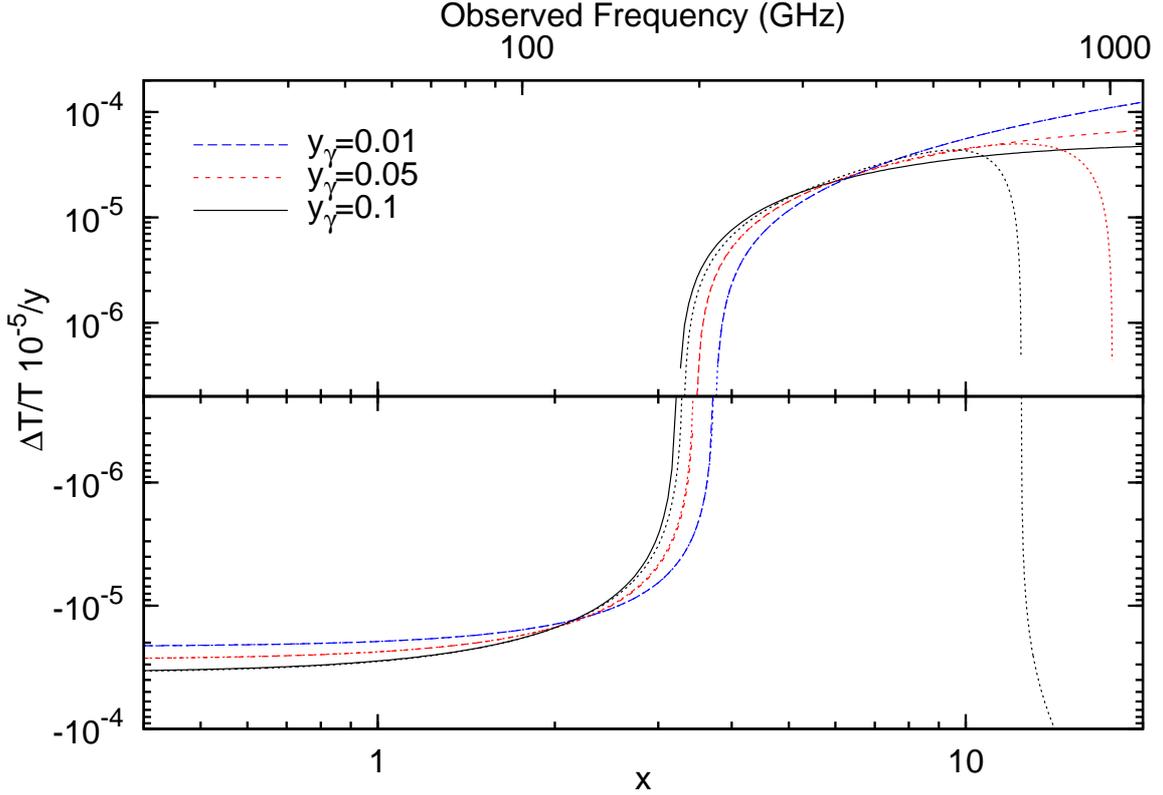}}
\caption{\label{anfig} Analytic solution given by Eq. \eqref{aneqcubic}
  including terms up to $\yg^3$ is shown
 by dotted lines for $\yg=0.01,0.05,0.1$ from top to bottom (in the
 Rayleigh-Jeans and Wien tails) respectively. Numerical solutions are as
 marked. Fractional difference in the effective temperature, Eq. \eqref{teffeq},
 is plotted. The two solutions match well  for $\yg\ll1$ and $x\ll 1/\yg$.  At $\yg=0.1$ the
 analytic solutions has the correct approximate shape. Errors are plotted
 in Fig. \ref{anerrfig}}
\end{figure}

\begin{figure}
\resizebox{\hsize}{!}{\includegraphics{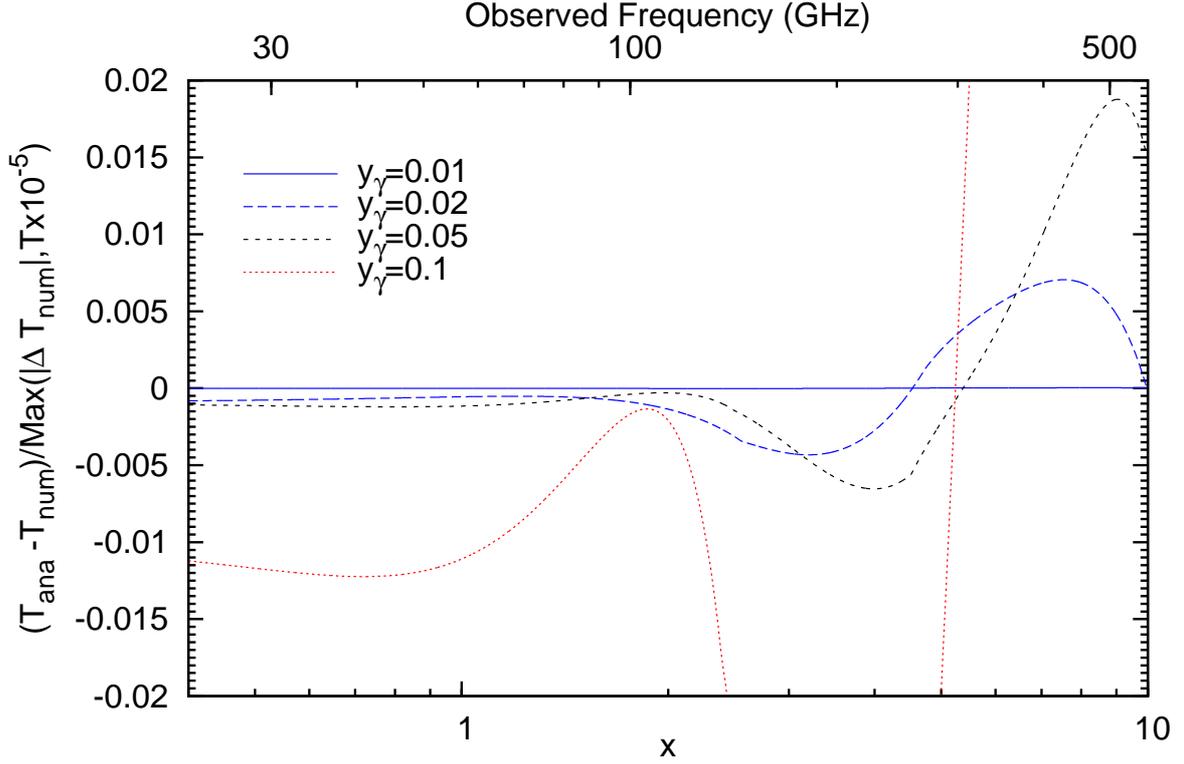}}
\caption{\label{anerrfig} Error in analytic solution defined by $\frac{\left(\Delta
  T/T_{\rm analytic}-\Delta T/T_{\rm numerical}\right)}{\max(|\Delta
  T/T|_{\rm numerical},10^{-5})}$. The analytic solution has better than $1\%$
accuracy over most of the frequency range of interest at $\yg\lesssim
0.05$. Accuracy deteriorates quickly at larger values of $\yg$.}
\end{figure}

%\begin{figure}
%\resizebox{\hsize}{!}{\includegraphics{analyticy1.eps}}
%\caption{\label{anfigy1} Analytic solutions for $\yg=0.1$ including successive
%terms in the Taylor series expansion are shown. Also shown is the numerical
%solution.}
%\end{figure}

\begin{figure}
\resizebox{\hsize}{!}{\includegraphics{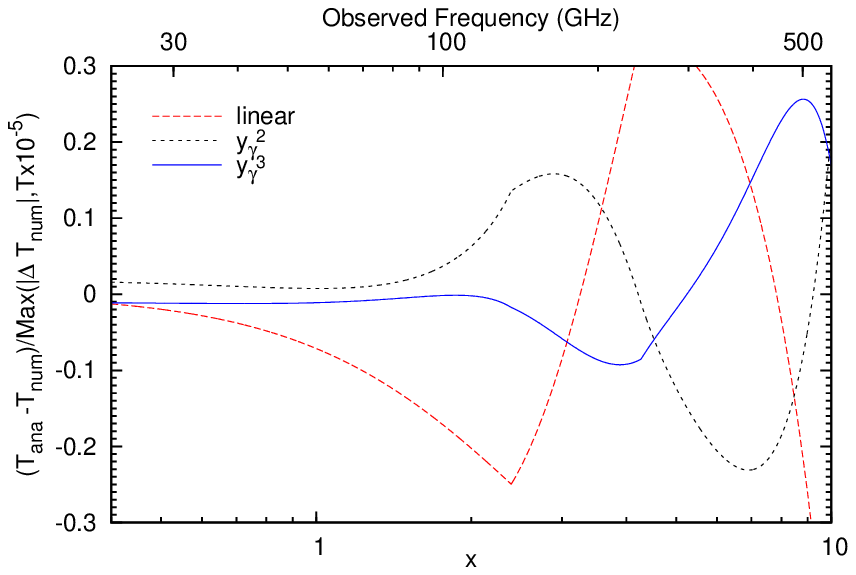}}
\caption{\label{anerrfigy1} Error in analytic solution defined by $\frac{\left(\Delta
  T/T_{\rm analytic}-\Delta T/T_{\rm numerical}\right)}{\max(|\Delta
  T/T|_{\rm numerical},10^{-5})}$ for $\yg=0.1$ for analytic solutions at
different orders. The analytic solution has better than $10\%$ at
$x\lesssim 6$. It can be seen that the convergence of the Taylor series is
very slow at $\yg=0.1$ with the analytic solution oscillating around the true
solution with inclusion of successive terms.}
\end{figure}
The solution to the Kompaneets equation for small distortions, with the initial spectrum being a
blackbody spectrum, for small values of $\yg$ parameter, $\yg\ll 1$, is given by the
Taylor  series Eq. \eqref{expan}, with the first two
coefficients/derivatives given by Eqs. \eqref{linsz} and
\eqref{szcorr}. The functions $f_{\y}(x)$ and $f_2(x)$ just involve
derivatives of the Planck spectrum and are easily calculated.
\begin{align}
f_{\y}(x)&=\frac{-2   e^x x}{\left(e^x-1\right)^5}
   \left[-x^2-7 x-8+e^{3 x} \left(x^2-7 x+8\right)\right.\nonumber\\
&\left.+e^x \left(-2 x^3-9 x^2+7 x+24\right)+e^{2 x} \left(-2 x^3+9 x^2+7 x-24\right)\right]\label{fyeq}\\
f_2(x)&=\frac{  e^x x}{\left(e^x-1\right)^5} \left[x^3+12 x^2+34 x+16+e^{3
    x} \left(x^3-12 x^2+34 x-16\right)\right.\nonumber\\
&\left.+e^{2 x} \left(11 x^3-36 x^2-34
    x+48\right)+e^x \left(11 x^3+36 x^2-34 x-48\right)\right]
\end{align}

For an initial $\y$-type spectrum, $n(x,0)=\npl(x)+\y n_{\y}(x)$, $\DTe(\yg=0)=5.4\y$,  we can similarly
expand the solution $n(x,\yg)$ in Taylor series, keeping only terms linear in
$\y$ since COBE/FIRAS \citep{cobe} already constrains the average
cosmological distortion.
$\y\lesssim 10^{-5}$, $\yg$ on the other hand covers a wide range and is
$>1$ at $z>1.45\times 10^5$). $\yg$ in the solutions refers to  the
total  $\yg$ integrated from  the time of energy injection to the time where we want
to calculate the final distortion.
\begin{align}
n(x,\yg)=\npl(x)+\y n_{\y}(x) +\yg\frac{\partial n}{\partial \yg}(x,0) + \frac{\yg^2}{2}\frac{\partial^2 n}{\partial \yg^2}(x,0)+\frac{\yg^3}{6}\frac{\partial^3 n}{\partial \yg^3}(x,0)+\mathcal{O}(\yg^4)
\end{align}
We can calculate the coefficients in Taylor series iteratively by using
Kompaneets equation. The first derivative is thus given by the Kompaneets
equation. Second derivative is obtained by differentiating Kompaneets
equation with respect to $\yg$ and using the solution of first derivative and
so on. Derivatives of $\DTe(\yg)$ are also easily obtained using
Eq. \eqref{te}. First two derivatives are $\id \DTe/\id \yg|_{\yg=0}
\approx -21.45 \y $, and $\id^2 \DTe/\id
 \yg^2|_{\yg=0} \approx 323.6 \y $.
Thus,
\begin{align}
n(x,\yg)&=\npl+ \y\left[n_{\y} +\yg \left(5.4 n_{\y} +
    f_{\y}\right)+\frac{\yg^2}{2}\left(-21.45n_{\y}+5.4f_{\y}+g_{\y}^{(2)}\right)\right.\nonumber\\
&\left.+\frac{\yg^3}{6}\left(323.6n_{\y}-21.45f_{\y}+5.4g_{\y}^{(2)}+g_{\y}^{(3)}\right)\right]
+O(\yg^4).\label{aneqcubic}
\end{align}
In general $m^{th}$ derivative is given by (for $m>1$),
\begin{align}
\frac{\partial^m n}{\partial \yg^m}(x,0)&=\sum_{i=0}^{m-1}\frac{1}{\y}\frac{\id^i \DTe}{\id \yg^i}g_{\y}^{(m-i-1)}+g_{\y}^{(m)},
\end{align}
where $g_{\y}^{(0)}=n_{\y}$, $g_{\y}^{(1)}=f_{\y}$ and for $m>0$ the $g_{\y}^{(m)}$ functions are given
recursively by 
\begin{align}
g_{\y}^{(m+1)}=\frac{1}{x^2}\frac{\partial }{\partial
  x}x^4\left(\frac{\partial g_{\y}^{(m)} }{\partial x} +g_{\y}^{(m)}(1+2\npl)\right).
\end{align}

 We compare the 
 the analytic solution including first three terms (up to order $\yg^3$)  and the numerical solution in
 Fig. \ref{anfig}. 
The two solutions match very well for $\yg\ll 1,x\ll 1/\yg$ and the error at $\yg=0.05$ is
 $\lesssim 1\%$ for $x\lesssim 7$. However, for larger values of $\yg$, the solution quickly
 deteriorates, and at $\yg=0.1$ the error is of order $10\%$ at $x\lesssim
 6$. For $\yg\lesssim 0.01$, the linear order term is enough to give $\sim
 1\%$ accuracy. Fig. \ref{anerrfigy1} compares the analytic solution including up to linear,
 quadratic and cubic terms with numerical solution for $\yg=0.1$. It can be
 seen that with the inclusion of  successive terms, the analytic solution
 oscillates around the true solution and convergence is quite slow. 
{Table \ref{tbl} gives the approximate maximum values of $\yg$ where the error at $x\lesssim 6$
is  below $1\%,5\%,10\%$ at linear, quadratic and cubic orders in $\yg$.
\begin{table}
\begin{tabular}{|c|c|c|c|}
\hline
 Error  & $1\%$ & $5\%$ & $10\%$ \\
\hline
linear &0.01  &0.03&0.04\\
quadratic&0.03 &0.06&0.08\\
cubic &0.05&0.08&0.1\\
\hline
\end{tabular}
\caption{\label{tbl}Upper bound on errors in the analytic solutions at
  different orders in $\yg$. 
    The upper bounds in different columns are  reached at the values of
    $\yg$ shown.  This
  table can be used as a rough guide when using analytic solutions.} 
\end{table}
}
\end{appendix}

\end{document}